\documentclass[twocolumn]{aastex701}

\usepackage{subcaption}
\usepackage{soul}
\usepackage{amssymb,microtype,siunitx,booktabs}
\begin{document}

\title{Spectral Disentangling Reveals Deep CNO-cycle Exposure in ET\,Cru}

\correspondingauthor{G\"{o}khan Y\"{u}cel}

\author[0000-0002-9846-3788]{G\"{o}khan Y\"{u}cel}
\affiliation{Department of Space Sciences and Technologies, Faculty of Science, Akdeniz University, 07058, Antalya, T\"{u}rkiye}\email[show]{gokhannyucel@gmail.com}  

\author[0000-0002-3125-9010]{Volkan Bak{\i}\c{s}}
\affiliation{Department of Space Sciences and Technologies, Faculty of Science, Akdeniz University, 07058, Antalya, T\"{u}rkiye}
\affiliation{Graduate Institute of Natural and Applied Sciences, Akdeniz University, 07058, Antalya, T\"{u}rkiye}
\email{vbakis@akdeniz.edu.tr}

\author[0000-0003-4752-4365]{Christian Nitschelm}
\affiliation{Centro de Astronomía (CITEVA), Universidad de Antofagasta, Avenida Angamos 601, Antofagasta 1270300, Chile}
\email{avc@a.com}

\author[0000-0002-0296-233X]{Timur \c{S}ahin}
\affiliation{Department of Space Sciences and Technologies, Faculty of Science, Akdeniz University, 07058, Antalya, T\"{u}rkiye}
\affiliation{Graduate Institute of Natural and Applied Sciences, Akdeniz University, 07058, Antalya, T\"{u}rkiye}
\email{timursahin@akdeniz.edu.tr}

\author[0000-0003-3884-974X]{Ferhat G\"{u}ney}
\affiliation{Department of Space Sciences and Technologies, Faculty of Science, Akdeniz University, 07058, Antalya, T\"{u}rkiye}
\affiliation{Graduate Institute of Natural and Applied Sciences, Akdeniz University, 07058, Antalya, T\"{u}rkiye}
\email{ferhatguneyau@gmail.com}

\begin{abstract}

Binary stars undergoing mass transfer provide unique laboratories for testing stellar evolution. 
Here, we present a comprehensive photometric and spectroscopic analysis of the semi-detached system ET\,Cru. 
Using spectral disentangling, we independently determined the effective temperatures and chemical abundances of both components with high precision, including nine elements (eleven species). 
We find masses of $13.41\,M_\odot$ and $6.00\,M_\odot$ for the primary and secondary, respectively, with uncertainties of only $\sim$1.3\%. 
The radii are $5.58\,R_\odot$ and $5.68\,R_\odot$, measured to within 0.4\% and 0.5\%. 
Surface gravities are constrained to better than 1\%, while effective temperatures are determined to within 3--5\%. 
The secondary exhibits extreme chemical anomalies, with severe carbon depletion and nitrogen enrichment far exceeding those reported in classical Algol systems. 
Multi-wavelength spectral energy distribution modelling yields a distance of $\sim$2.5\,kpc, inconsistent with the \textit{Gaia} DR3 parallax, suggesting systematic astrometric uncertainties in the parallax distance. 
Together, these results establish ET\,Cru as a benchmark Algol-type binary, revealing direct spectroscopic evidence of deep CNO-cycle exposure in the donor and confirming the primary star as a rejuvenated gainer. 
ET\,Cru thus provides a chemically and dynamically illustrative case for understanding advanced binary interactions and the late evolutionary stages of massive-star evolution.

\end{abstract}



%------------------------------------------------------------------

\section{Introduction}
\label{sec:int}

Stellar astrophysics and evolution studies are fundamentally constrained by the accuracy of fundamental stellar parameters. Eclipsing binary systems serve as critical laboratories in this endeavor, as their unique geometry enables the direct and model-independent determination of properties such as mass and radius with exceptional precision, often better than 1\% \citep{Andersen1991, Torres2010, Serenelli2021}. This precision, further enhanced by space-based photometry \citep{Southworth2021}, underpins the vital role of eclipsing binaries in calibrating stellar models. The ensemble properties of eclipsing binaries have subsequently fueled extensive theoretical investigations and empirical relations, profoundly advancing our understanding of stellar structure and evolution \citep[e.g.][]{Malkov2003, malkov2007, Gafeira2012, Eker2015, Eker2018, Eker2020, Eker2024, Eker2025}.

Spectral analysis of eclipsing binary systems is inherently challenging, as the observed spectrum represents the composite light of both stellar components. One approach to address this problem is to construct synthetic spectra for each component and combine them to reproduce the observed spectrum \citep{Aschenbrenner2024}. However, modern techniques allow this difficulty to be overcome by disentangling the individual component spectra directly from the composite observations \citep{Hadrava1995, Ilijic2001}. This has enabled researchers to analyze each component of a binary system individually, thereby allowing reliable determination of fundamental parameters such as effective temperature and chemical abundances \citep[e.g.][]{Hensberge2000, Pavlovski2005, Kolbas2014, Kolbas2015, Dervisoglu2018, Pavlovski2023, Kovalev2024, Yucel2025, Yucel2026b}.

The evolution of eclipsing binaries is different from that of single stars because the gravitational attraction of the component stars causes mass transfer from one star to another over time. These evolved systems are called semi-detached binary systems. The intriguing part of these systems is that, when investigated, in most of them, the secondary, less massive one is the donor star. This paradox was first investigated by \cite{Crawford1955}, and later, a well-accustomed solution was provided by several researchers \citep{Hilditch2001}. These features of semi-detached binary systems make them well-studied astronomical objects, providing accurate fundamental parameters of component stars, as well as the evolutionary statuses and mass transfer properties of the stars. Despite the large sample size of such system studies, there is still a need for analysis of eclipsing binary systems, especially those that include massive components \citep{Malkov2020}. Despite extensive studies on eclipsing binaries, individual high-mass systems remain underexplored. In this context, ET\,Cru emerges as a particularly valuable target, offering an opportunity to investigate its fundamental parameters and chemical properties in detail.

\begin{figure*}
    \centering
    % İlk şekil
    \begin{subfigure}[t]{0.30\linewidth}
        \centering
        \includegraphics[width=\linewidth]{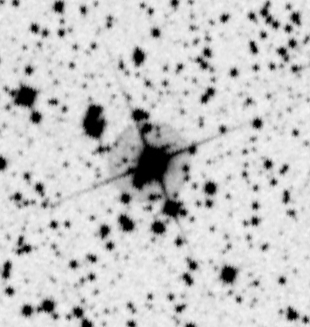}
        \caption{The \textit{DSS2} $B$-band field of view around ET\,Cru.}
        \label{fig:DSS2-field}
    \end{subfigure}
    \hfill
    % İkinci şekil
    \begin{subfigure}[t]{0.37\linewidth}
        \centering
        \includegraphics[width=\linewidth]{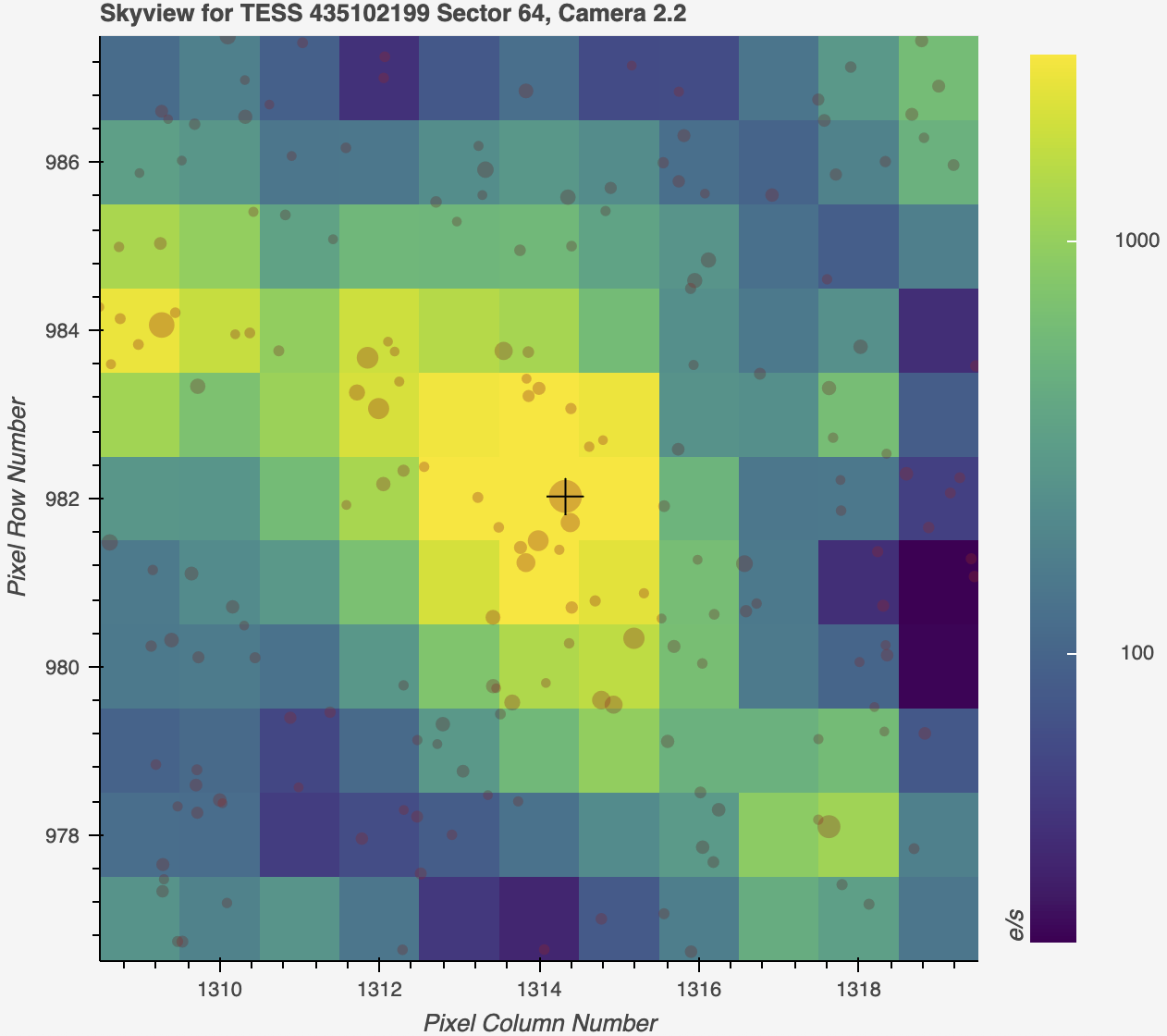}
        \caption{The \textit{TESS} field of view around ET\,Cru in Sector 64.}
        \label{fig:tess-field}
    \end{subfigure}
    \hfill
    % Üçüncü şekil
    \begin{subfigure}[t]{0.30\linewidth}
        \centering
        \includegraphics[width=\linewidth]{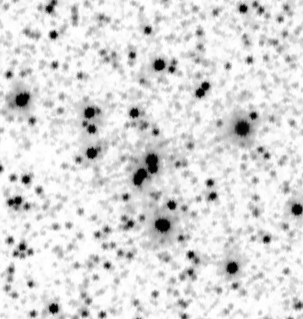}
        \caption{The \textit{2MASS} combined $JHK$-band field of view around ET\,Cru.}
        \label{fig:2mass-field}
    \end{subfigure}
    \caption{Field images of ET\,Cru from short to long wavelengths: DSS2 Blue (optical), \textit{TESS} (red-optical), and \textit{2MASS} (near-infrared). This sequence highlights the appearance of the system across different wavelength regimes. The dimension of images is $3.85\times3.85\,\rm{arcmin}^2$.}
    \label{fig:fields-comparison}
\end{figure*}

ET\,Cru (HD 111505, TYC 8988-1466-1; $l=312^{\rm o}\!\!.81296$, $b=+2^{\rm o}\!\!.20796$) was discovered as a variable star by \citet{Strohmeier1965}. The spectral class was determined as B2/3 III by \citet{Houk1975}. Its relatively short period was determined to be 2.04388 d by \citet{Otero2004}. \citet{Ijspeert2021} has reported it's a high-mass binary system by analyzing \textit{TESS} \citep{Ricker2015} data. There are no significant studies in the literature specific to ET\,Cru. In this study, we investigated ET\,Cru by combining high-resolution spectroscopic data with high-precision photometric data, and obtained its fundamental parameters. The chemical properties of the system are also discussed. Thus, the present study provides the first detailed spectroscopic and photometric analysis of ET\,Cru, contributing to a broader understanding of massive semi-detached binaries.

In this study, we present the first comprehensive analysis of ET\,Cru, determining its fundamental parameters to $<$1-3\% precision and deriving the atmospheric chemical abundances of both components to trace the system's mass-transfer history.

The remainder of this paper is organized as follows. Section 2 describes the photometric (ASAS, \textit{TESS}, and \textit{Gaia}) and spectroscopic (FEROS) observations and the initial measurements of the radial velocities. Section 3 details the simultaneous analysis of the radial-velocity and light-curve data to determine the orbital and fundamental parameters of the system. Section 4 presents the application of spectral disentangling to isolate the component spectra. Section 5 outlines the methodology for atmospheric parameter determination and detailed chemical abundance analysis. Section 6 discusses the derived astrophysical parameters, distance estimates, and significant chemical patterns. Finally, Section 7 summarizes our conclusions and places ET\,Cru in the context of the Algol binary evolution.

%------------------------------------------------------------------
\section{Data}
\subsection{Photometric Data}

Photometric observations of ET\,Cru are available from multiple surveys, including ASAS, \textit{TESS}, and \textit{Gaia}.

The ASAS-3 (All Sky Automated Survey; \citealt{Pojmanski1997}) observed ET\,Cru between December 2002 and July 2009 in the Johnson $V$ band with 60\,s exposures. ASAS provides photometry measured with five different apertures (MAG0--MAG4), corresponding to radii of 2 to 6 pixels. Considering the relatively crowded field of ET\,Cru (Figure~\ref{fig:fields-comparison}), we adopted the MAG1 (3-pixel) data to minimize blending effects. Only data points with the highest quality flag (“A”) were retained, resulting in 659 measurements used in the analysis.

\textit{TESS} observed ET\,Cru in five sectors (11, 37, 38, 64, and 65) between April 2019 and June 2023, with exposure times ranging from 1800\,s to 200\,s. The light curves were obtained from the MIT Quick Look Pipeline \citep[QLP;][]{Huang2020a, Huang2020b, Kunimoto2021, Kunimoto2022}. The TESS light curves for individual sectors are shown in Figure~\ref{fig:TESS}. Only data points passing the \texttt{quality\_mask=hard} criterion were considered. Details of the \textit{TESS} mission are given by \citet{Ricker2015}. The use of TESS data in the modelling is discussed in Section~3.3.

\textit{Gaia} \citep{Gaia2023} observed ET\,Cru between October 2014 and May 2017, providing photometric measurements in three bands ($G$, $G_{\rm BP}$, and $G_{\rm RP}$), with approximately 60 datapoints per band.

\subsection{Spectroscopic Data}

Spectroscopic data of ET\,Cru were obtained using the Fiber-fed Extended Range Optical Spectrograph (FEROS; \citealt{Kaufer1999}), mounted on the MPG/ESO 2.2-m telescope at La Silla Observatory, Chile. FEROS provides a resolving power of $R \approx 48\,000$ and covers the spectral range from 3500 to 9200\,\AA. Observations were conducted over five nights, resulting in a total of twelve spectra. The spectra were reduced using the standard FEROS pipeline, which includes bias subtraction, flat-field correction, and wavelength calibration.

%------------------------------------------------------------------
\section{Analysis}
\subsection{Updated Linear Ephemeris}

The TESS observations of the ET~Cru system span more than two years. As discussed in Section~2.1.1, although nearby stars are present within the TESS aperture, the times of minima can be determined with high precision, allowing the derivation of reliable light elements. Within this framework, we measured a total of 111 times of minima (56 primary and 55 secondary) from the TESS photometric light curves. Together with one additional time of minimum measurement from ASAS data \citep{Pojmanski1997}, reported by \citet{Otero2004}, the total number of times of minima used in this study is 112.

Using the TESS data, we constructed the $O-C$ diagram and performed a linear fit, yielding the updated light elements given below:

\[
Min\,I\,(HJD) = 2459321.8934(2) + E\times 2.04386211(39)
\]

where the numbers in parentheses indicate the uncertainty in the last quoted digit.

\begin{figure}
    \centering
    \includegraphics[width=0.9\linewidth]{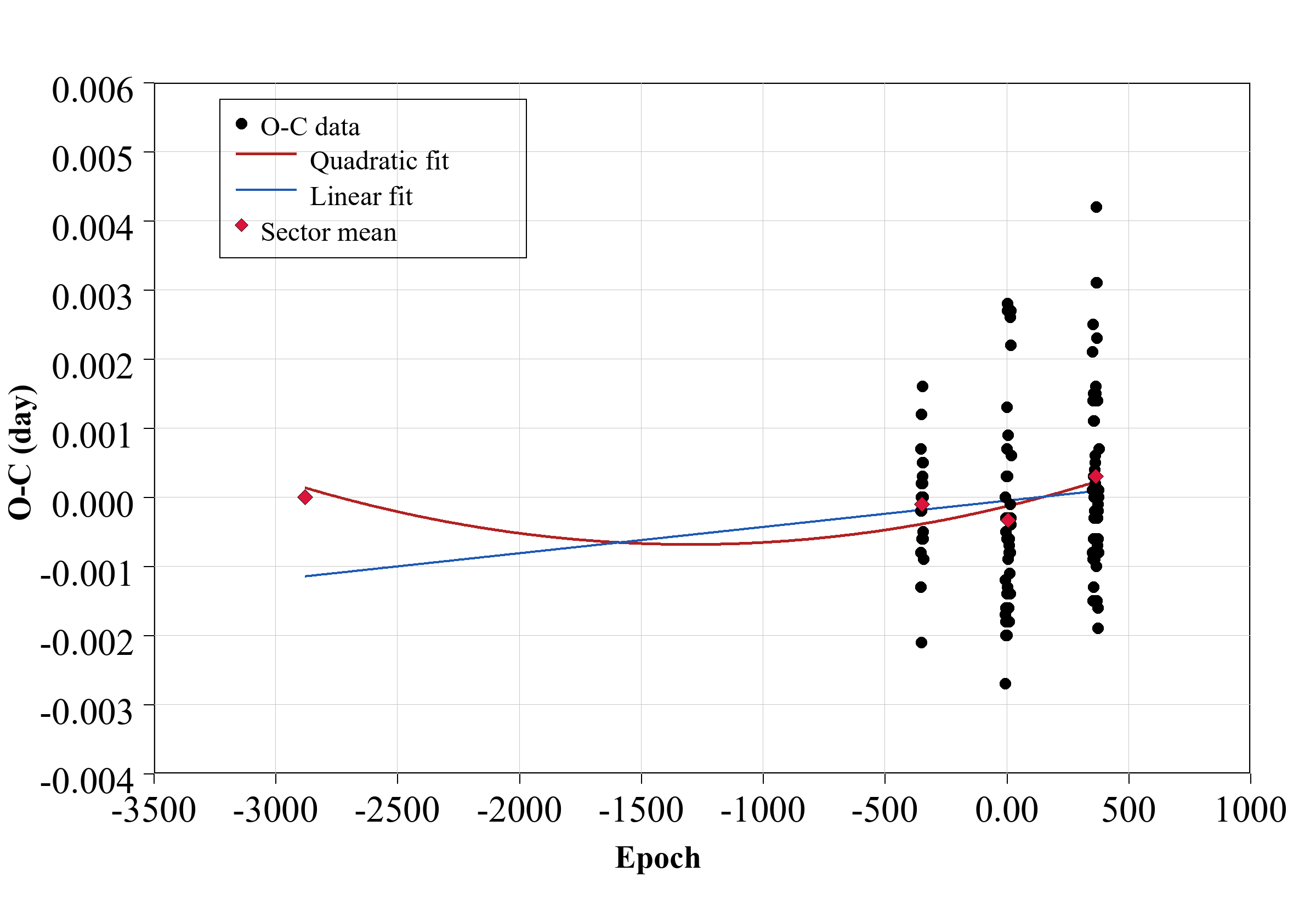}
    \caption{Linear and quadratic fits to the $O-C$ residuals. The far left point belongs to ASAS, while the rest of the data is from the TESS Sectors. Sector mean values are shown with a red circle.}
    \label{fig:oc}
\end{figure}

In addition to the linear representation, we also applied a quadratic fit to the $O-C$ residuals, including the ASAS measurement (see Fig.~\ref{fig:oc}), in order to examine whether the orbital-period variation can be described more accurately by a secular trend.

A comparison of the fit statistics shows that the quadratic model provides a slightly better representation of the data than the linear one. The linear fit yields an RMSE of $1.2792\times10^{-3}$ day, an AIC value of $-1501.5$, and a BIC value of $-1496.1$. For the quadratic fit, the RMSE decreases to $1.2667\times10^{-3}$ day, while the AIC and BIC values become $-1501.7$ and $-1493.5$, respectively. The lower RMSE and AIC values indicate that the quadratic model is statistically preferred, although the improvement over the linear fit remains modest.

The positive quadratic term suggests a possible long-term increase in the orbital period of the system. Using the quadratic coefficient, the period change rate was derived as
\[
{\dot{P}=(3.2 \pm 2.1)\times10^{-10}\ {\rm day\,day^{-1}},}
\]
{or equivalently}
\[
{\dot{P}=(1.2 \pm 0.8)\times10^{-7}\ {\rm day\,yr^{-1}}.}
\]

Assuming conservative mass transfer, this period increase, combined with the component masses and orbital period of the system listed in Table~\ref{tab:parameters}, corresponds to a mass transfer rate of
\[
{\dot{M}=(2.1 \pm 1.4)\times10^{-7}\ M_{\odot}\,{\rm yr^{-1}}.}
\]

Therefore, although the linear ephemeris remains adequate as a first-order description of the currently available timing data, the quadratic fit appears to account better for the observed $O-C$ behaviour and may indicate the presence of a secular increase in the orbital period of ET~Cru.

\subsection{Measurements of Radial Velocities}

Radial velocity (RV) measurements of the components of ET\,Cru were performed following the methodology described by \citet{Yucel2025}. Briefly, the procedure involved several steps. First, spectral regions that were largely unaffected by interstellar extinction lines were selected. Next, a Nonlocal Thermodynamic Equilibrium (NLTE) synthetic spectra grid was constructed, considering the spectral type of the system, as reported by SIMBAD. The grid was generated using \texttt{iSpec} \citep{Blanco2014, Blanco2019} in combination with \texttt{SYNSPEC} \citep{Hubeny1995, Hubeny2017}. The observed spectra were then compared simultaneously with the synthetic spectra to derive the RVs for both components. The HJD-corrected RVs and their weighted average errors are listed in Table\ref{tab:rv}.

\begin{table}
    \centering
        \setlength{\tabcolsep}{4pt}
    \renewcommand{\arraystretch}{0.95}
    \caption{Radial velocities: HJD is heliocentric Julian date, RV1 the radial velocities of the primary component, and RV2 the radial velocities of the secondary. One sigma errors are given for both radial velocity columns.} 
    \label{tab:rv}
    {\footnotesize
    \begin{tabular}{lcrrrr}
    \hline
        HJD & Phase & RV1 & Error & RV2 & Error  \\
        (day) & ($\phi$) & \multicolumn{4}{c}{(km\,s$^{-1}$)} \\
        \hline
        2455647.64139 & 0.305 & -130.35 & 3.20 & 264.57  & 1.94 \\ 
        2455647.64972 & 0.309 & -126.71 & 2.30 & 261.59  & 1.32 \\ 
        2455648.68013 & 0.813 & 121.79  & 2.56 & -264.65 & 5.16 \\ 
        2455648.69173 & 0.819 & 118.05  & 3.81 & -258.36 & 4.48 \\ 
        2455649.55586 & 0.241 & -129.74 & 3.69 & 279.96  & 1.92 \\ 
        2455649.56767 & 0.247 & -134.44 & 3.38 & 282.13  & 2.14 \\ 
        2455650.51993 & 0.713 & 119.32  & 2.51 & -281.57 & 4.74 \\ 
        2455650.53209 & 0.719 & 124.80  & 2.57 & -278.92 & 6.62 \\ 
        2455651.56537 & 0.225 & -132.68 & 3.22 & 274.44  & 2.24 \\ 
        2455651.57509 & 0.229 & -129.32 & 5.14 & 277.46  & 3.05 \\ 
        2455651.68013 & 0.281 & -138.22 & 4.91 & 264.03  & 4.11 \\ 
        2455651.69551 & 0.288 & -132.82 & 4.75 & 268.09  & 2.42 \\ \hline

    \end{tabular}
    }
\end{table}

\begin{figure}
    \centering
    \includegraphics[width=0.9\linewidth]{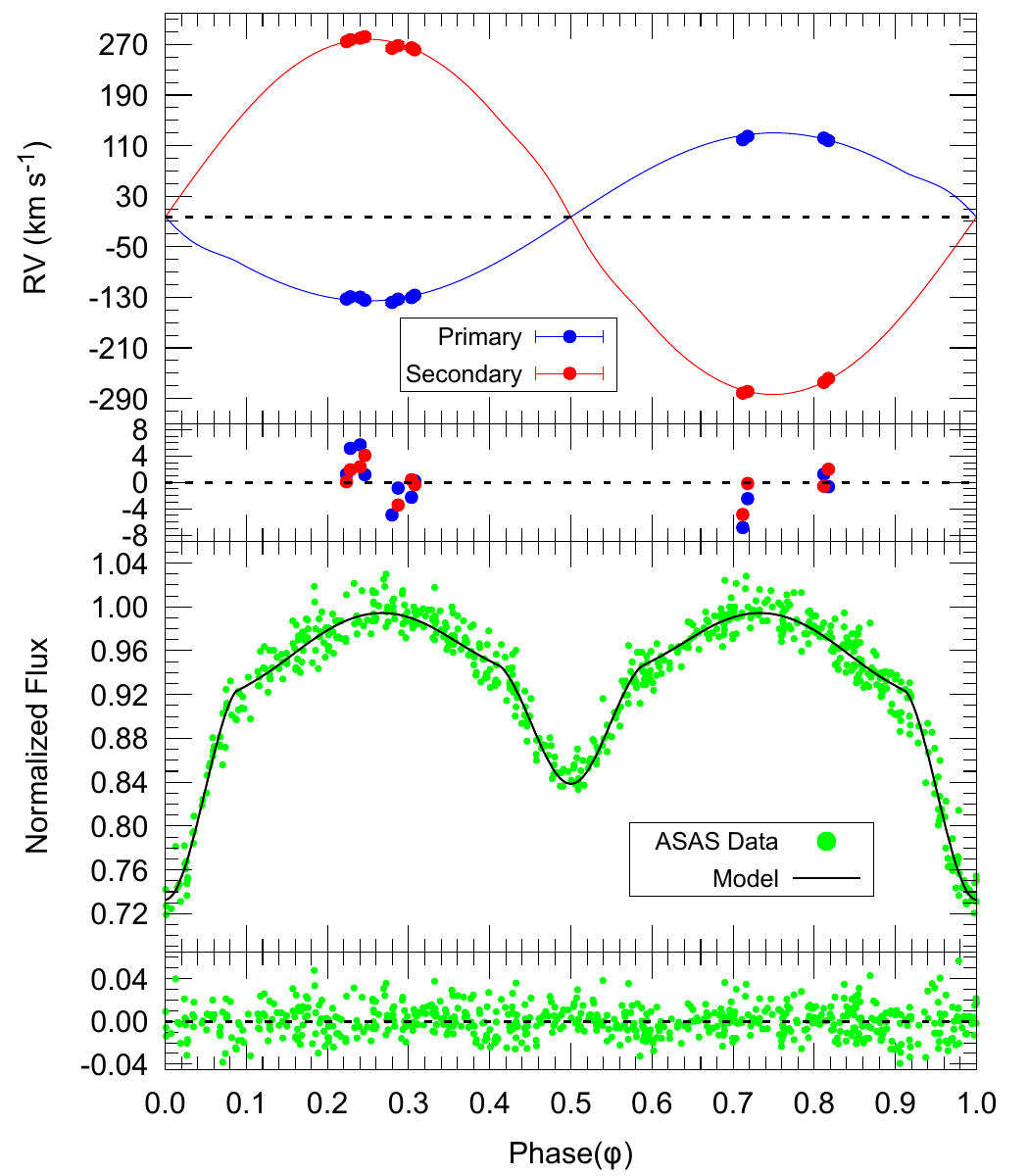}
    \caption{Observed RVs and photometric light curves of ET\,Cru, together with the best-fitting models and corresponding residuals. The blue and red filled circles indicate the RV measurements of the primary and secondary components, respectively. In the LC panel, green dots show the ASAS photometric data, and the black curve represents the best-fitting LC solution.}
    \label{fig:rvlc}
\end{figure}

\begin{figure*}
    \centering
    \includegraphics[width=0.49\linewidth]{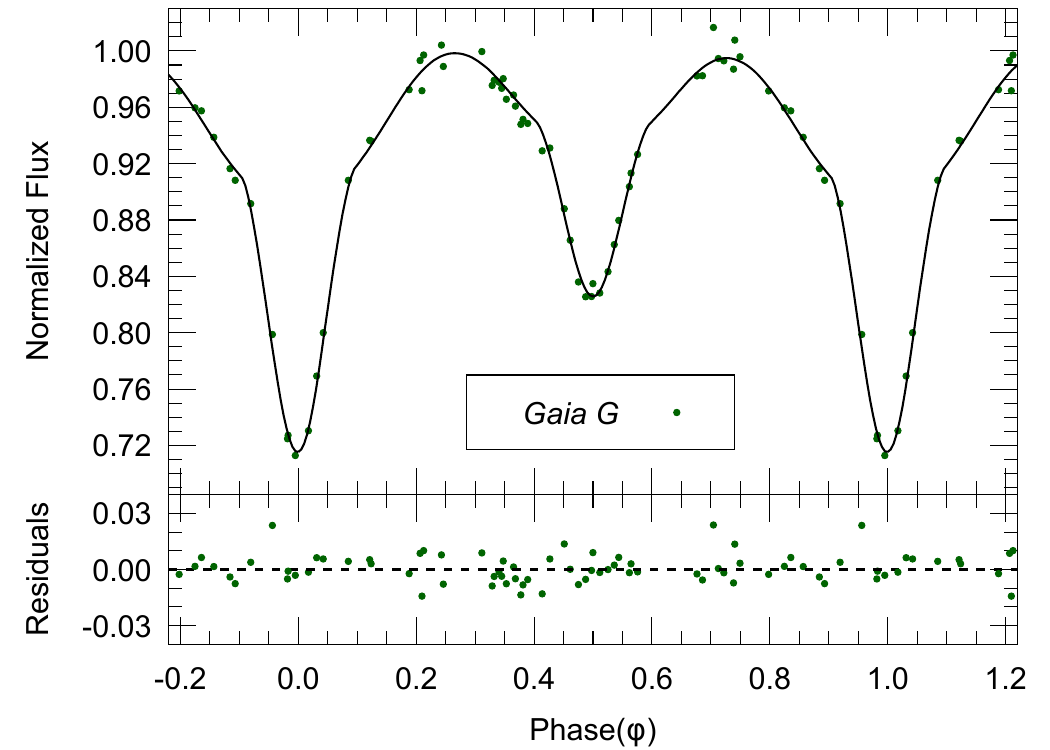}
    \includegraphics[width=0.49\linewidth]{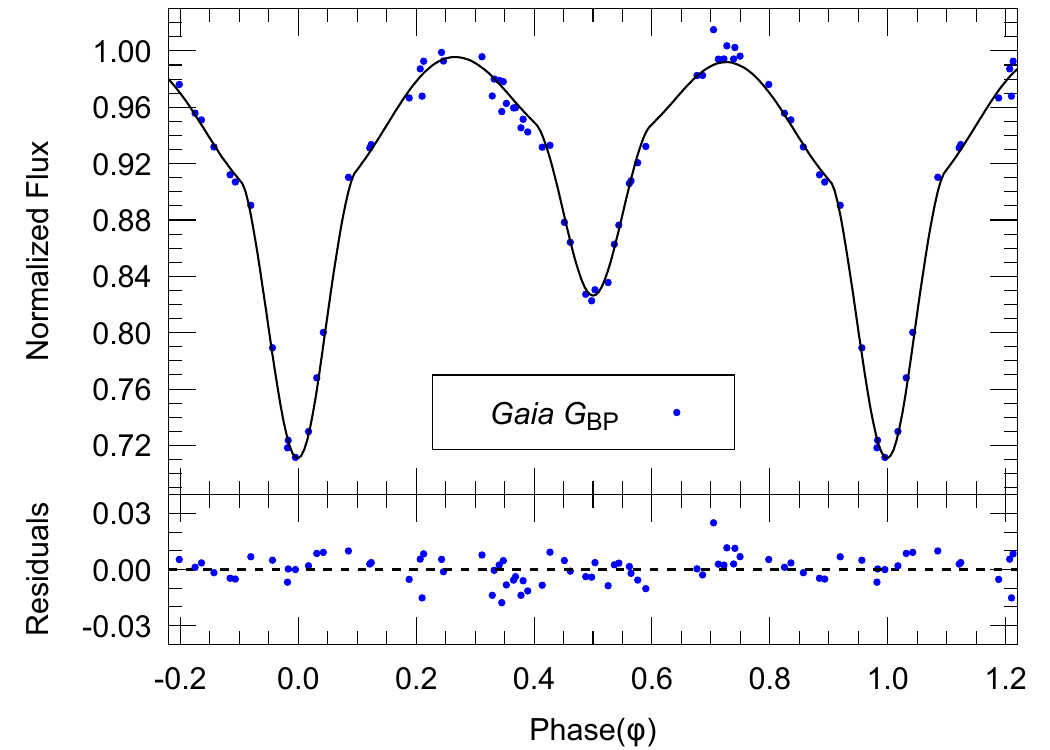} \\
    \includegraphics[width=0.49\linewidth]{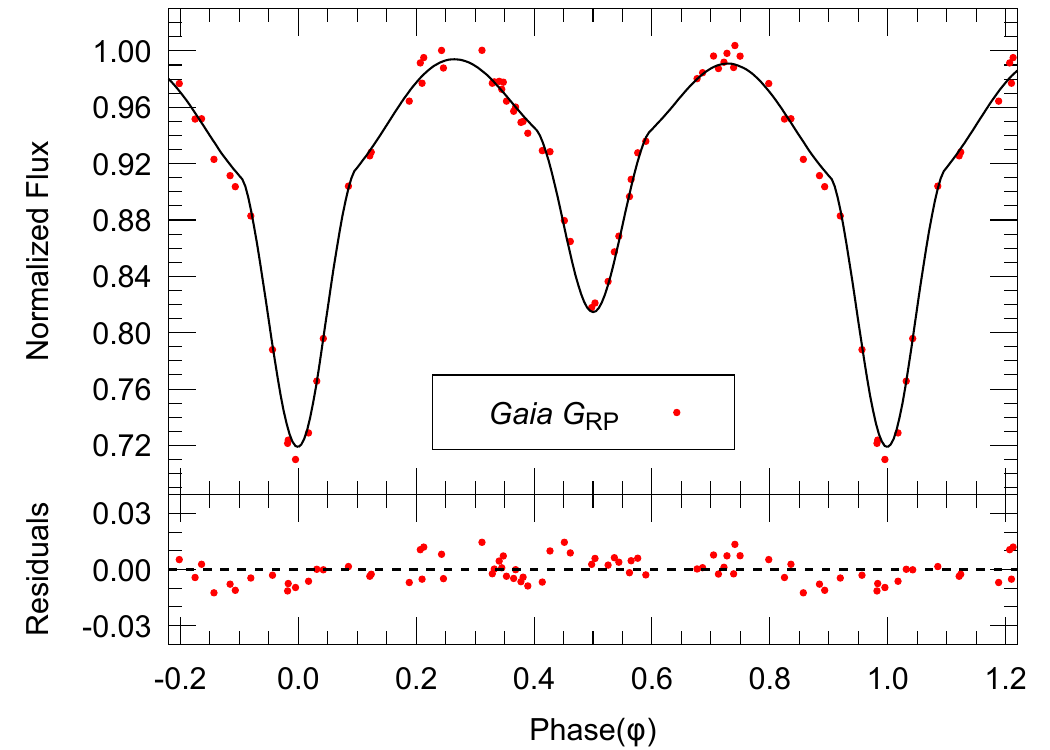}
    \includegraphics[width=0.49\linewidth]{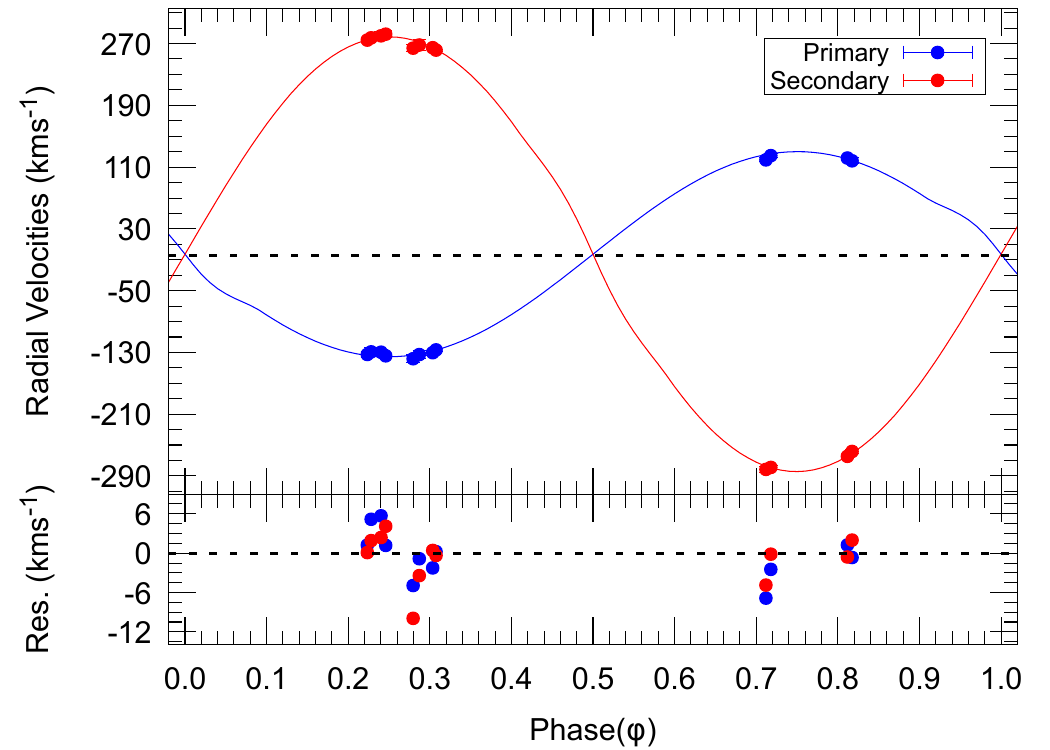}
    \caption{Observations and best fitting LC and spectroscopic orbital models for \textit{Gaia} LCs and FEROS RV data, respectively.}
    \label{fig:gaia_lcs}
\end{figure*}

\subsection{Analysis of RV and Light Curve (LC)} \label{sect:RV_LC_analysis}

We employed the PHysics Of Eclipsing BinariEs code (\texttt{PHOEBE} v2.4; \citealt{phoebe2_2016, phoebe2_2018, phoebe2_2020, phoebe2_2020b}) to determine the fundamental parameters of ET\,Cru by modelling its RVs and light curves. Because no previous studies have established the system morphology, we explored both detached and semi-detached configurations guided by the LC shape. The analysis converged to a semi-detached configuration, which provides the best representation of the observed data.

The resulting solutions are consistent with those obtained from earlier implementations, demonstrating that the inferred system parameters are robust against differences in the numerical implementation.

In addition to ASAS photometry, we included \textit{Gaia} DR3 light curves in the $G$, $BP$, and $RP$ bands in our modelling. The ASAS $V$-band light curve and the \textit{Gaia} $G$, $BP$, and $RP$ light curves were modelled independently, and separate solutions were derived for the ASAS and \textit{Gaia} data sets. These data provide an independent constraint on the system variability and yield a period fully consistent with the adopted value. The solutions obtained from the \textit{Gaia} passbands are mutually consistent and are also in good agreement with the solution derived from the ASAS $V$-band data, thereby supporting the robustness of the inferred system parameters. Owing to the higher quality and multi-band nature of the \textit{Gaia} photometry, we adopt the \textit{Gaia}-based solution throughout the remainder of this paper, unless explicitly stated otherwise.

We also examined the available TESS photometric data. The TESS light curves exhibit significant amplitude variations between different sectors and exposure times, affecting both maxima and eclipse depths. While variations at maxima may be attributed to physical effects such as stellar activity, the changes in eclipse depths are not expected to depend on sector or cadence, and instead point to the presence of systematic effects related to data reduction or normalization. 

In particular, the photometric amplitudes derived from TESS are significantly larger than those obtained from both ASAS and \textit{Gaia} data, and cannot be consistently reproduced within a common modelling framework without introducing additional scaling. Therefore, TESS data were not included in the quantitative modelling. For completeness, the TESS light curves for different sectors and cadences are presented and discussed in the Appendix.

During modeling, the conjunction time ($T_0$), orbital period ($P$), and temperature of the primary component were fixed. The primary temperature was set to 23\,000 K, according to \citet{Houk1975} as a starting point but was later refined spectroscopically (see \S\ref{Models}). The following parameters were adjusted: mass ratio ($q$), semi-major axis ($a$), systemic velocity ($v_\gamma$), orbital inclination ($i$), temperature ratio ($T_2/T_1$), potential of the primary component ($\Omega_1$),  monochromatic luminosity of the primary ($L_1$) , and the contribution of the third light as a fraction ($L_3$). The gravity-darkening exponents and albedos of both components were set to 1.0 \citep{Rafert1980}, while the limb-darkening coefficients were internally calculated by \texttt{PHOEBE} using the stellar atmosphere models of \citet{Castelli2003}.

\begin{figure*}
    \centering
    \includegraphics[width=0.9\linewidth]{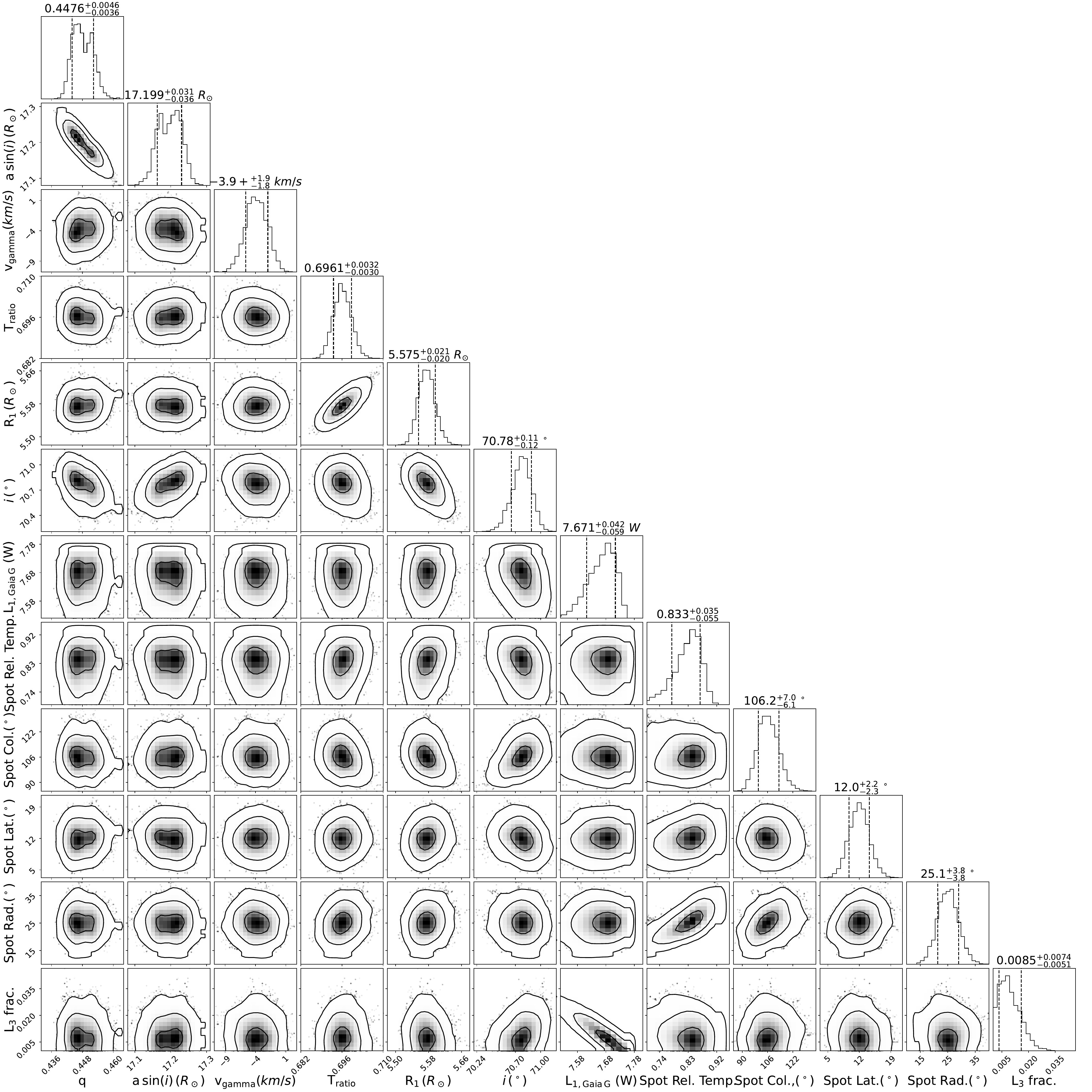}
    \caption{The corner plot of the posteriors for the fundamental parameters of the components of ET\,Cru from $Gaia\,G$ LC.}
    \label{fig:mcmc}
\end{figure*}

To refine the stellar parameters and obtain robust uncertainty estimates, we performed a Markov Chain Monte Carlo (MCMC) analysis that built-in feature of \texttt{PHOEBE} \citep{Foreman2013}. MCMC sampling was used to explore parameter correlations and provide statistically robust uncertainties. The computation employed 128 walkers, each evolved for 1000 iterations to ensure adequate sampling of the posterior distributions. The LC and spectroscopic orbit models are presented in Figures~\ref{fig:rvlc} and \ref{fig:gaia_lcs}, while the fundamental parameters and their heuristic errors are shown in Figure~\ref{fig:mcmc} and listed in Table~\ref{tab:solution}.

In the \textit{Gaia} light curves of ET\,Cru, a noticeable flux decrease is observed at the ingress of the primary eclipse. Such behaviour is commonly seen in mass-transferring Algol-type systems and has been attributed to the presence of gas streams originating from the secondary component (e.g. \citealt{rara2016, hhcar2021}). In this scenario, the transferred material can partially obscure the primary star just before the onset of the primary minimum, leading to an additional attenuation of the observed flux.

\begin{figure*}
    \centering
    \includegraphics[width=0.32\linewidth]{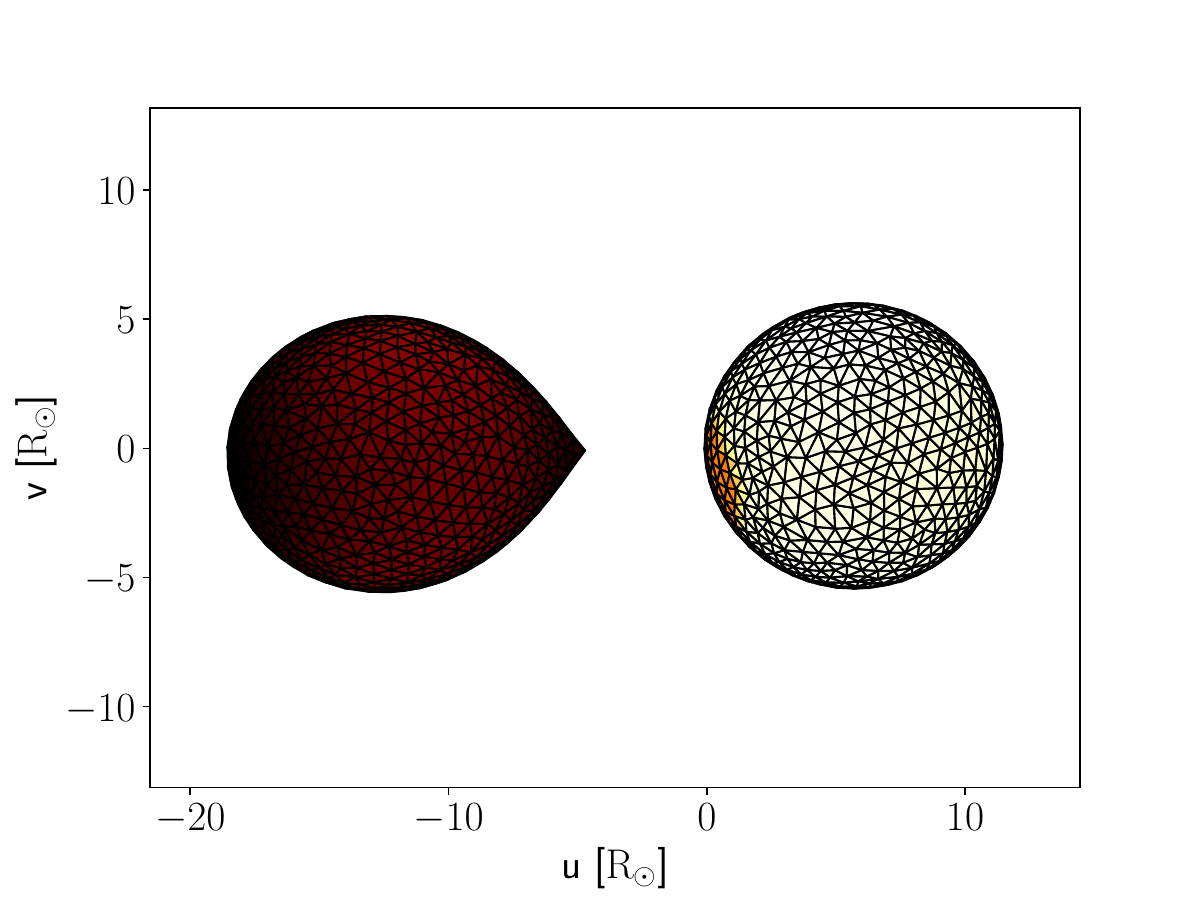}
    \includegraphics[width=0.32\linewidth]{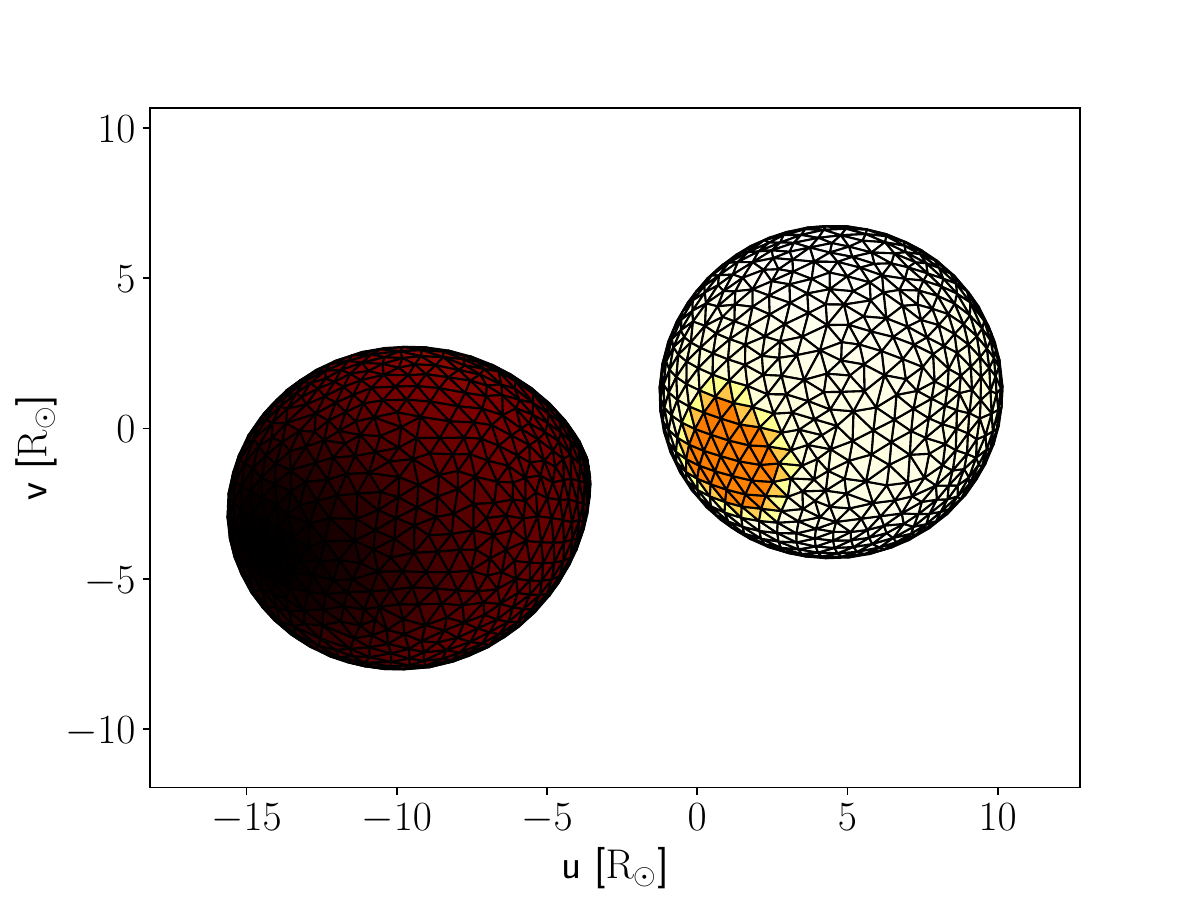}
    \includegraphics[width=0.32\linewidth]{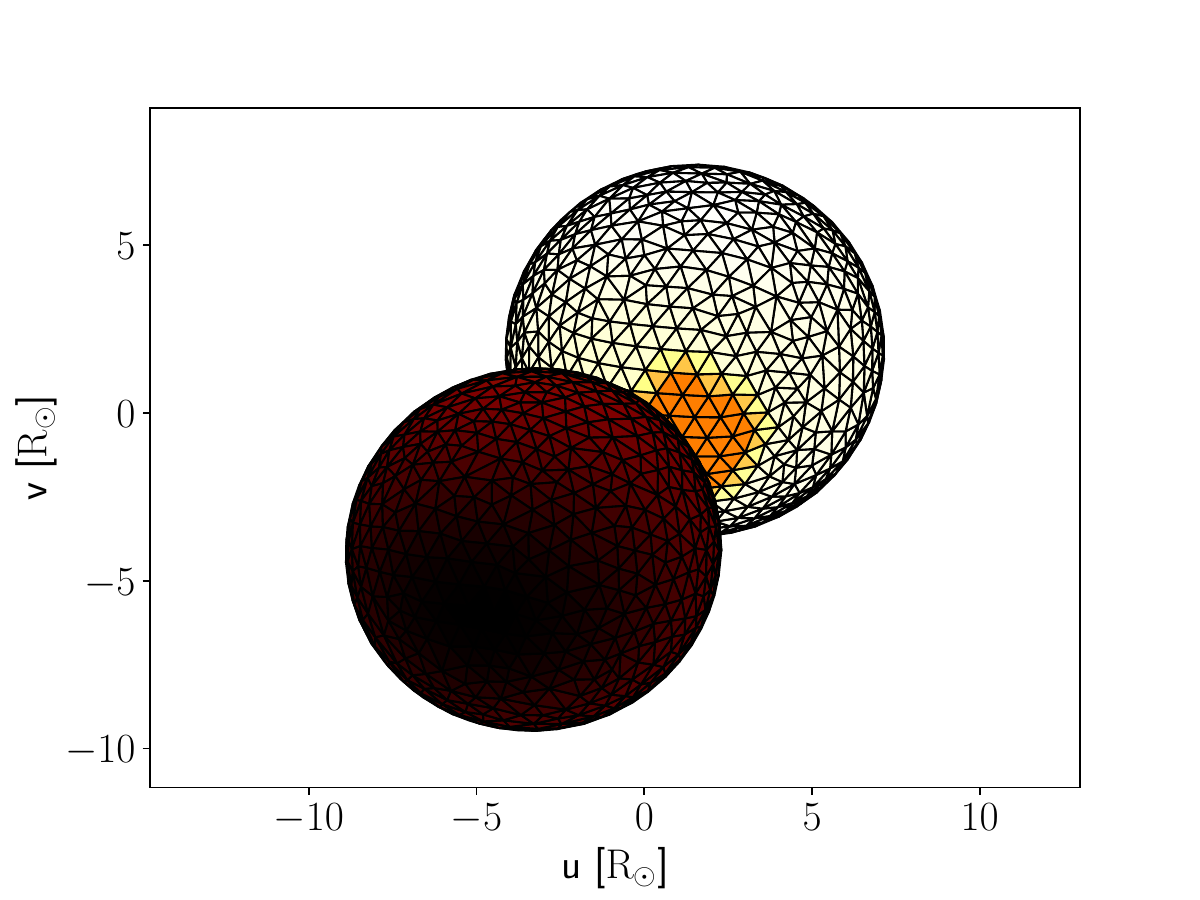}
    \caption{Mesh plots of the binary system ET\,Cru at orbital phases 0.75 (left), 0.85 (middle), and 0.95 (right), generated with \texttt{PHOEBE} v2.4. The colour scale represents the local effective temperature across each stellar surface, while the mesh structure traces the distorted Roche-lobe-filling components. The sequence illustrates the changing aspect of the system as it approaches secondary eclipse (phase 0.75) and moves towards the quadrature (phase 0.85) and pre-eclipse (phase 0.95) configurations.}
    \label{fig:mesh}
\end{figure*}

A practical approach to account for this effect within the framework of light curve modelling is to introduce a localized cool spot on the surface of the primary component (see Figure~\ref{fig:mesh}), representing the impact region of the accretion flow. This method has been successfully applied in previous studies (e.g. \citealt{rara2016, hhcar2021}) and provides a reasonable approximation of the observed asymmetry in the light curve.

\begin{table*}
	
\centering
\caption{Best-fit physical parameters of the ET\,Cru binary system derived from the analysis of RV and LC data, using the ASAS--$V$ and $Gaia$ passbands separately. } 
\label{tab:absolutepar}
\begin{tabular}{lcccc}\toprule
    & \multicolumn{2}{c}{ASAS--$V$} & \multicolumn{2}{c}{$Gaia$} \\
\midrule
Parameter   &  Primary  & Secondary  &  Primary  & Secondary  \\
\midrule
Separation ($a$ ($R_\odot$)) & \multicolumn{2}{c}{$18.24_{-0.06}^{+0.07}$}    & \multicolumn{2}{c}{$18.214_{-0.050}^{+0.046}$}       \\
Mass ratio     (\emph{q})                              & \multicolumn{2}{c}{$0.460_{-0.004}^{+0.004}$}  & \multicolumn{2}{c}{$0.4476_{-0.0036}^{+0.0046}$} \\
Systemic velocity (\(V_\gamma\) (km s$^{-1}$))  & \multicolumn{2}{c}{$-2.8_{-0.7}^{+0.8}$}        & \multicolumn{2}{c}{$-3.9_{-1.8}^{+1.9}$}   \\
Eccentricity    (\emph{e})                              & \multicolumn{4}{c}{0}                               \\
Orbital inclination (\emph{i} ($^\circ$))   & \multicolumn{2}{c}{$69.60_{-0.22}^{+0.21}$}  & \multicolumn{2}{c}{$70.78_{-0.12}^{+0.11}$}    \\
Temperature ratio  ($T_\mathrm{eff,b}/T_\mathrm{eff,a}$)   &  \multicolumn{2}{c}{$0.691_{-0.005}^{+0.005}$}  & \multicolumn{2}{c}{$0.6961_{-0.0030}^{+0.0032}$} \\
Mass (\emph{M} ($M_\odot$))   &  $13.342_{-0.176}^{+0.184}$  & $6.137_{-0.101}^{+0.105}$ & $13.408_{-0.153}^{+0.135}$  & $6.001_{-0.082}^{+0.089}$ \\
Radius (\emph{R} ($R_\odot$))  &  $5.297_{-0.151}^{+0.159}$  & $5.723_{-0.040}^{+0.033}$  &  $5.575_{-0.021}^{+0.020}$  & $5.676_{-0.027}^{+0.030}$ \\
Surface gravity ($\log g$ (cgs)) &  $4.115_{-0.031}^{+0.031}$  & $3.711_{-0.012}^{+0.013}$   &  $4.073  _{-0.008}^{+0.008}$  & $3.708_{-0.010}^{+0.011}$ \\
Light ratio  ($l/l_\mathrm{total}$) &  $0.595_{-0.010}^{+0.012}$  & $0.404_{-0.012}^{+0.010}$  & $0.6108_{-0.0005}^{+0.0006}$  & $0.3807_{-0.0006}^{+0.0005}$\\
3rd light ratio  ($l_3/l_\mathrm{total}$) &  \multicolumn{2}{c}{$0.001_{-0.001}^{+0.001}$}  & \multicolumn{2}{c}{$0.0085_{-0.0051}^{+0.0074}$}\\
\bottomrule
\label{tab:solution}
\end{tabular}
\end{table*}

\subsection{Spectral Disentangling}

In binary systems, the observed spectrum represents a composite of both stellar components, complicating reliable parameter estimation if analyzed without specialized techniques. One approach to infer the effective temperatures and chemical compositions of the components is to construct synthetic spectra for each star and combine them according to their respective light ratios, enabling a direct comparison with the observations \citep{Tomasella2008, Yucel2022}. However, a more established method involves isolating the individual component spectra from the composite data by applying specialized algorithms and subsequently analyzing each spectrum independently. This procedure, known as spectral disentangling \citep{Simon1994}, has been successfully employed in numerous studies of binary systems \citep[see for recent applications][]{Skarka2025, Bakis2025}. 

\begin{figure*}
    \centering
    \includegraphics[width=0.99\linewidth]{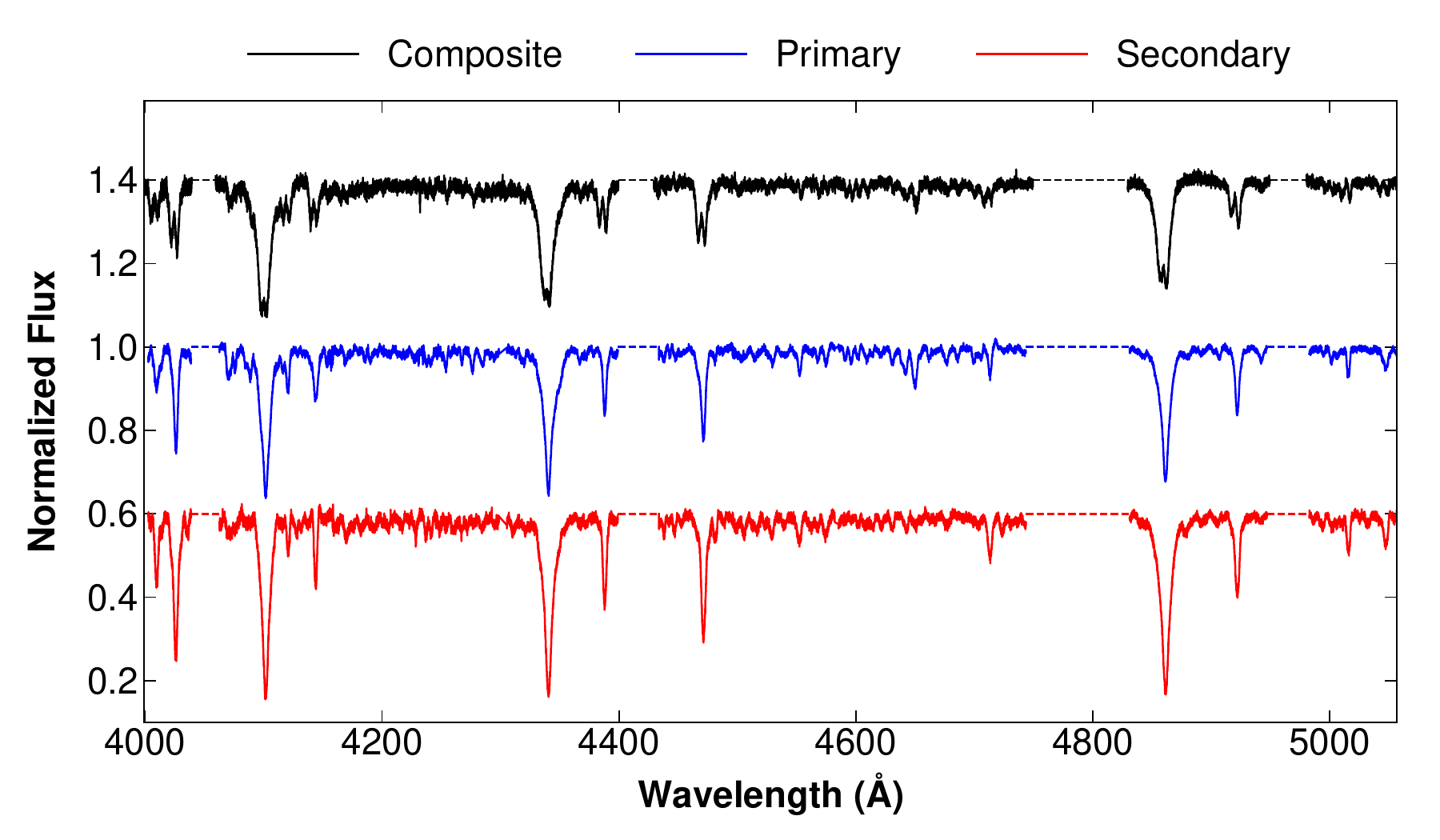}
    \caption{The observed composite spectrum, $\phi=0.819$, of the semi-detached binary ET\,Cru (black) and its decomposition into the individual contributions of the primary (blue) and secondary (red) components in the 4000–5000 Å region. This disentangling, essential for isolating the spectral features of each star, provided the foundation for the subsequent determination of orbital dynamics, atmospheric parameters, and detailed chemical abundances.}
    \label{fig:composite}
\end{figure*}

In the present work, spectral disentangling was performed using the \textsc{korel} code \citep{Hadrava1995}, which implements a Fourier-based method that simultaneously solves for the orbital parameters and separates the component spectra. This approach ensures that the blended spectral features do not bias the determination of the stellar parameters. The observed spectra were divided into multiple regions, each containing 4096 bins and corresponding to wavelength intervals shorter than 50~\AA. During disentangling with \textsc{korel}, the spectral orbit of the system was solved simultaneously. In this study, while isolating the spectra of the individual components, we adopted the orbital parameters obtained in \S\ref{sect:RV_LC_analysis}. The disentangled spectra of the components are shown in Figure~\ref{fig:composite}.

%------------------------------------------------

\subsection{Model Atmosphere Parameters and Abundances} \label{Models}

The spectroscopic analysis presented in this work is based on a recently introduced comprehensive atomic line list designed for hot B-type stars \citep{Guney2026}. This list incorporates transitions from 13 key photospheric elements (He, C, N, O, Ne, Mg, Al, Si, P, S, Cl, Ar, and Fe), with detailed line counts provided in the source publication. It was previously benchmarked using the standard candidate star HR~1765 \citep{csahin2019high} and the Algol-type system u~Her \citep{Kolbas2014}. Here, we apply and further validate this line list in the analysis of the rapidly rotating Algol binary, ET\,Cru.

A reliable abundance analysis requires precise atmospheric parameters, primarily the effective temperature (\(T_{\mathrm{eff}}\)) and surface gravity (\(\log g\)). For ET\,Cru, we resolved the inherent spectroscopic degeneracy between these parameters by leveraging the binary nature of the system. This process began with the application of spectral disentangling to the observed composite spectra, a technique that successfully separates the light from the two components. The result of this decomposition in the 4000--5000\,\AA\ region is presented in Figure~\ref{fig:composite}, which clearly shows the individual contributions of the primary and secondary stars to the total light. Although this technique recovers individual spectral line profiles, the absolute continuum levels for each component remain unspecified.

Therefore, determining the precise relative light contribution (\(l\)) of each star is essential to correctly renormalize the disentangled spectra, as illustrated in Figure~\ref{fig:composite}, and recover accurate line depths for the subsequent abundance analysis. We derived the necessary light ratios (\(l / (l_p + l_s)\)) from the solution of the \textit{Gaia G} photometric light curve. The \textit{Gaia G} passband fully encompasses the spectroscopic wavelength range, ensuring the validity of the photometric ratios for our analysis. The semi-detached configuration of ET\,Cru provides strong geometric constraints, allowing for precise determination of the stellar radii. By combining the photometric orbital solution with the dynamical solution from spectral disentangling, we independently calculated the masses and radii of both components. This yielded direct, model-independent values for \(\log g\) (see Table~\ref{tab:absolutepar}).

Fixing \(\log g\) to these independently determined values breaks the \(T_{\mathrm{eff}}\)--\(\log g\) degeneracy. Consequently, the profiles of the hydrogen Balmer lines can be used exclusively to constrain \(T_{\mathrm{eff}}\), free from the ambiguities introduced by correlated adjustments in electron density. This approach provides a robust and transparent determination of fundamental atmospheric parameters. 

To determine the parameter space for global $\chi^2$ minimization, we used the three hydrogen Balmer lines (H$\beta$, H$\gamma$, and H$\delta$) and Helium lines \citep[see][]{Guney2026} present in 4000 - 5056 \AA\, spectral range of the FEROS spectrum. These were fitted against an extensive library of pre-computed synthetic spectra from the BSTAR2006 grid\footnote{\url{https://tlusty.oca.eu/Tlusty2002/Tlusty-frames-BS06.html}}. This grid provides 1540 metal line-blanketed, NLTE, plane-parallel, and hydrostatic model atmospheres suitable for B-type stars. It samples 16 effective temperatures from 15,000 to 30,000 K (in steps of 1000 K), 13 surface gravities from $\log g = 1.75$ to 4.75 (step 0.25 dex), six chemical compositions (from metal-rich to metal-free), a solar helium abundance (He/H = 0.1 by number), and a microturbulent velocity of 2 km s$^{-1}$.

Both individual and combined fits of the Balmer lines were performed to derive $T_{\mathrm{eff}}$ and $\log g$. This analysis yielded the fundamental atmospheric parameters listed in Table~\ref{tab:etcru_params} for this system.

\begin{table}
\centering
\caption{Fundamental atmospheric parameters of the primary and secondary components of ET\,Cru.}
\label{tab:etcru_params}
\begin{tabular}{l cc}
\hline
\hline
Parameter & Primary & Secondary \\
\hline
$T_{\mathrm{eff}}$ (K)          & 29000 $\pm $1000 & 20200 $\pm $1000 \\
$\log g$ (cgs)                 & 4.15 $\pm$ 0.15  & 3.70 $\pm$ 0.30  \\
$[Fe/H]$ (dex)                   & 0.20 $\pm$ 0.15  & 0.20 $\pm$ 0.10  \\
$\xi$ (km\,s$^{-1}$) & 6.0 $\pm$ 0.5   & 6.0 $\pm$ 0.5   \\
$v\sin i$ (km\,s$^{-1}$)       & 145.0 $\pm$ 12.0 & 150.0 $\pm$ 2.0 \\
\hline
\end{tabular}
\end{table}

\begin{figure}[t]
    \centering
    \begin{minipage}{0.52\columnwidth}
        \centering
        \includegraphics[
            width=\linewidth,
            height=0.22\textheight,
            keepaspectratio,
            trim={3mm 2mm 3mm 2mm},clip
        ]{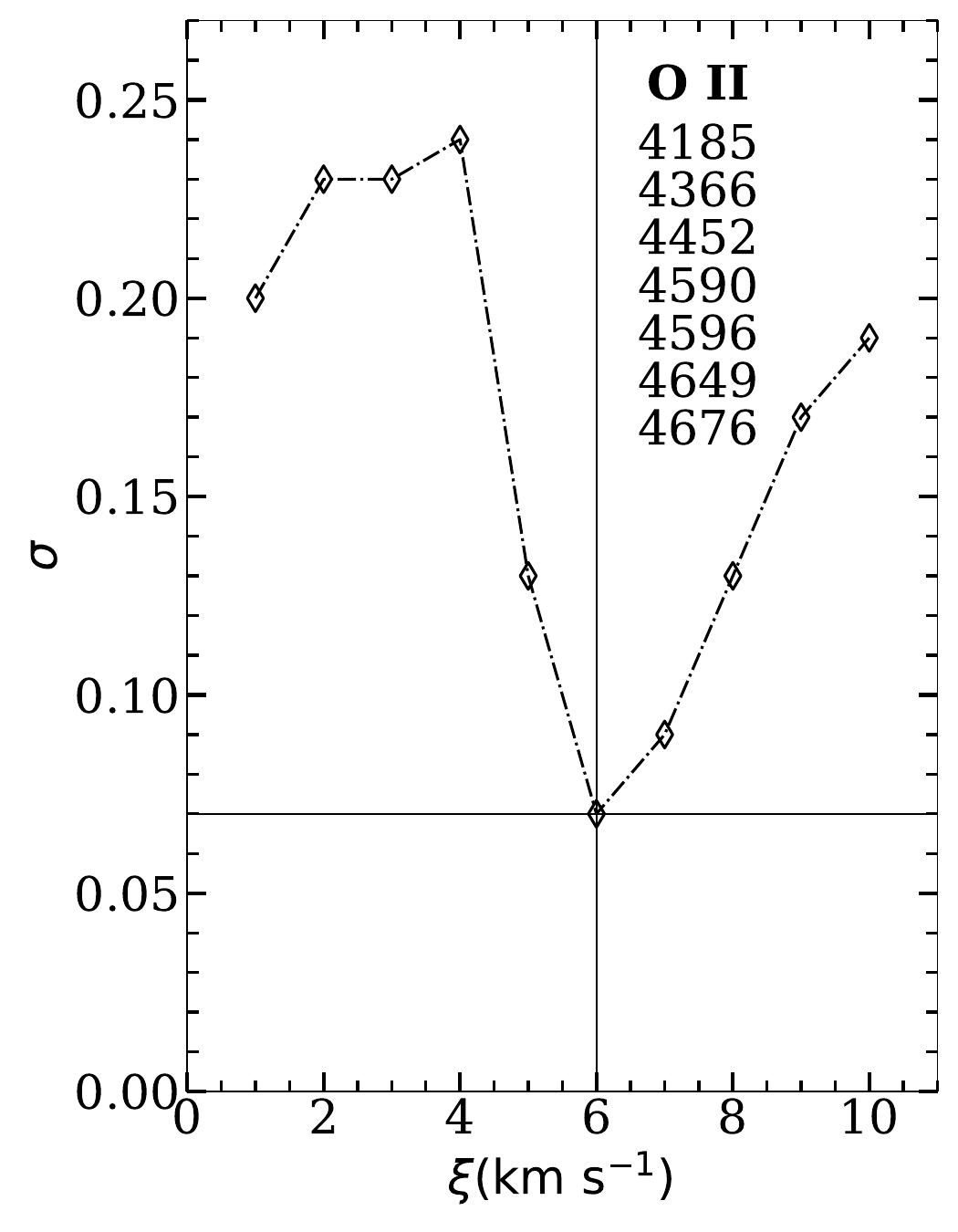}
    \end{minipage}\hfill
    \begin{minipage}{0.48\columnwidth}
        \centering
        \includegraphics[
            width=\linewidth,
            height=0.22\textheight,
            keepaspectratio,
            trim={3mm 2mm 3mm 2mm},clip
        ]{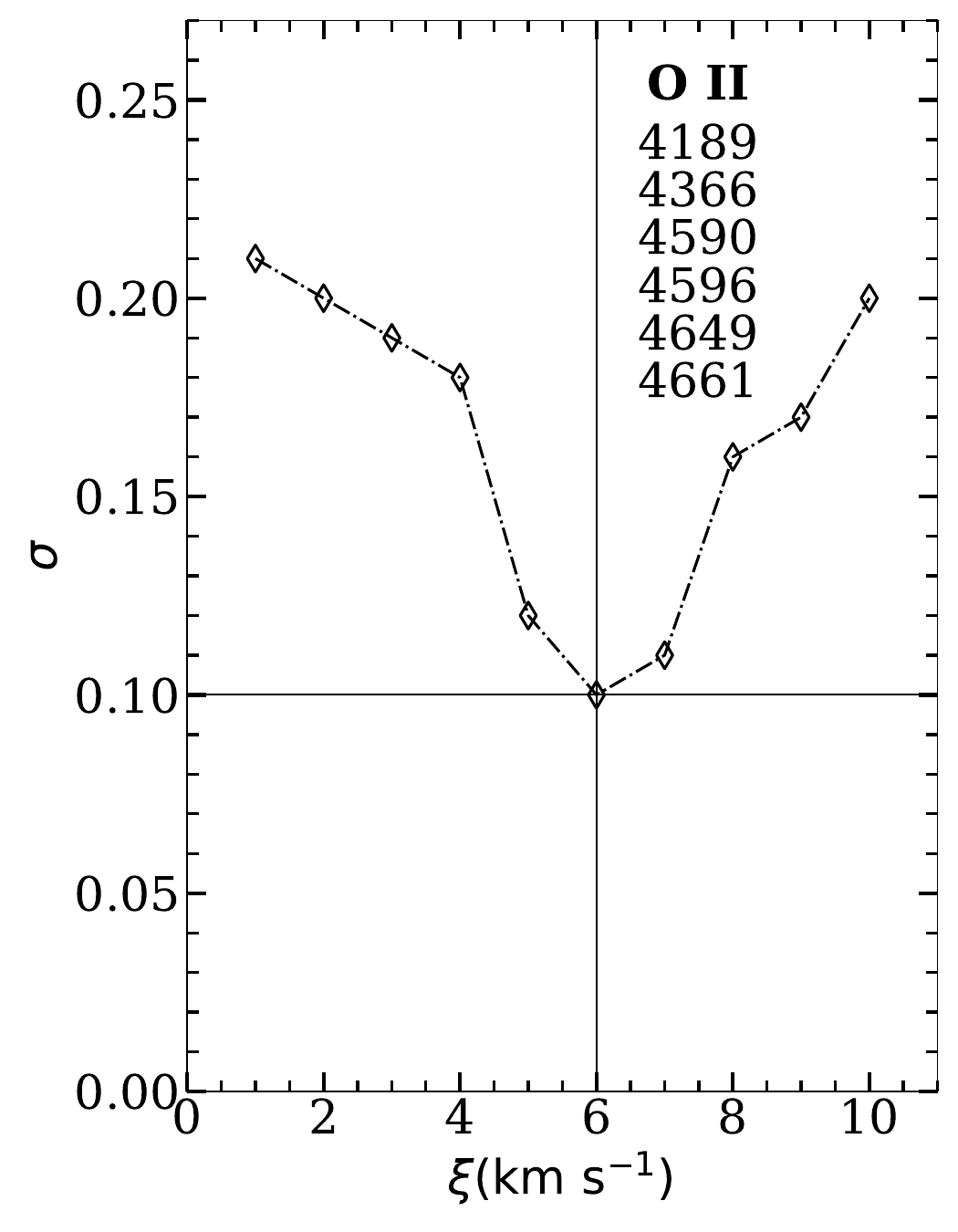}
    \end{minipage}
    \vspace{-4pt}
    \caption{Determination of the microturbulent velocity ($\xi$) for the primary (left panel) and secondary (right panel) components of the binary system ET\,Cru. The standard deviation ($\sigma$) of derived abundances from a set of selected O\,{\sc ii} lines is plotted against $\xi$. The adopted value for $\xi$ (indicated by the vertical and horizontal lines) corresponds to the point of minimum $\sigma$ for each component.}
    \vspace{-6pt}
    \label{fig:vmictest}
\end{figure}

Atmospheric modeling and spectral synthesis were performed using a well-established methodology for hot stars. We employed the \texttt{TLUSTY}\footnote{\url{https://tlusty.oca.eu/tlusty/index.html}} model atmosphere code \citep[][tlusty198]{Hubeny1995}, the \texttt{SYNSPEC} spectrum synthesis program \citep{Hubeny1995}, and the BSTAR2006 NLTE model grids \citep{lanz2007grid}.

A consistent treatment of spectral line broadening was performed. Stark broadening for the hydrogen Balmer lines was computed using the unified theory profiles implemented in \texttt{SYNSPEC} \citep{vidal1973hydrogen}, while the treatment for higher series members followed \citet{hubeny1994nlte}. For helium, we utilized broadening tables from \citet{barnard1974broadening} for the 4471\,\AA\ line, \citet{shamey1969stark} for the 4026, 4387, and 4922\,\AA\ lines, and \citet{dimitrijevic1984stark} for all other He\,{\sc i} lines. Metal lines were synthesized as Voigt profiles, with radiative, van der Waals, and Stark broadening included self-consistently using validated constants. The microturbulent velocity ($\xi$) for each component of ET\,Cru was determined spectroscopically using the standard method of minimizing the abundance scatter from weak metallic lines. For each component, we computed the standard deviation ($\sigma$) of the non-LTE oxygen abundances derived from a set of selected O\,{\sc ii} lines across a range of trial $\xi$ values. The adopted microturbulence, presented in Figure \ref{fig:vmictest}, corresponds to the $\xi$ value at which $\sigma$ reaches its minimum, ensuring that the derived abundances are independent of line strength.

\begin{figure*}
    \centering
    \includegraphics[width=0.99\linewidth]{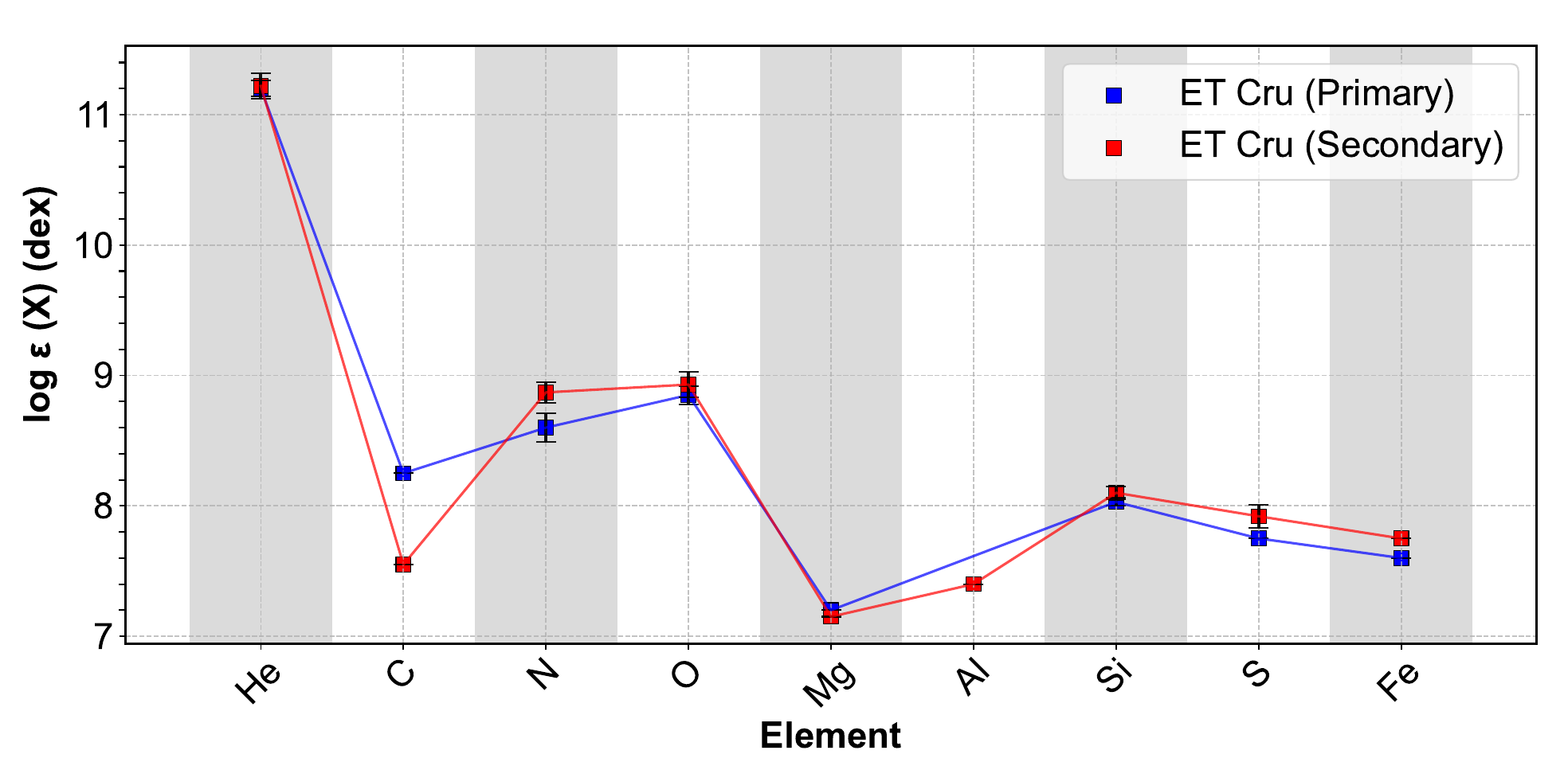}
    \caption{Comparison of photospheric elemental abundances (\(\log \varepsilon(X)\)) for the primary (gainer) and secondary (donor) components of the Algol-type binary system ET\,Cru. The plot shows the spectroscopically derived abundances of He, C, N, O, Mg, Al, Si, S, and Fe in both stars. Error bars represent \(1\sigma\) uncertainties.}
    \label{fig:abund_pattern}
\end{figure*}

The \texttt{TLUSTY} models were calculated in NLTE, incorporating the opacity from the bound-free and bound-bound transitions of H, He, C, N, O, Ne, Mg, Si, S, and Fe. The model atoms were comprehensive, including, for example, H\,{\sc i} (9 levels), He\,{\sc i} (24), C\,{\sc ii} (22), C\,{\sc iii} (46), N\,{\sc ii} (42), O\,{\sc ii} (48), Mg\,{\sc ii} (25), Si\,{\sc ii/iii/iv} (40, 30, 23), S\,{\sc ii/iii/iv} (33, 41, 38), and Fe\,{\sc ii/iii/iv} (36, 50, 43). Our analysis focused on weak and medium-strength metal lines for the abundance measurements.

\begin{table}[ht]
\centering
\caption{Elemental abundances (\(\log \epsilon(X)\)) and number of lines used (N) for the primary and
secondary components of ET\,Cru.}
\label{tab:etcru_abundances}
\begin{tabular}{l|cc|cc}
\hline \hline
\textbf{Element} & \multicolumn{2}{c|}{\textbf{Primary}} & \multicolumn{2}{c}{\textbf{Secondary}} \\
& \(\log \epsilon(X)\)(dex) & N & \(\log \epsilon(X)\)(dex) & N \\
\hline
He \textsc{i}  & \(11.20 \pm 0.06\) & 7 & \(11.22 \pm 0.11\) & 9 \\
C \textsc{ii}  & \(8.25\) & 1 & \(7.55\) & 2 \\
N \textsc{ii}  & \(8.60 \pm 0.11\) & 8 & \(8.87 \pm 0.08\) & 12 \\
O \textsc{ii}  & \(8.85 \pm 0.07\) & 11 & \(8.93 \pm 0.13\) & 6 \\
Mg \textsc{ii} & \(7.20\) & 1 & \(7.15\) & 1 \\
Al \textsc{iii}& --       & -- & \(7.40\) & 1 \\
S \textsc{ii}  & \(7.75\) & 1 & \(7.93 \pm 0.13\) & 3 \\
S \textsc{iii} & \(7.75\) & 2 & \(7.90 \pm 0.05\) & 2 \\
Si \textsc{ii} & \-- & -- & \(8.05\) & 1 \\
Si \textsc{iii}& \(8.03 \pm 0.03\) & 3 & \(8.12 \pm 0.05\) & 3 \\
Fe \textsc{iii}& \(7.60\) & 1 & \(7.75\) & 1 \\
\hline \hline
\end{tabular}
\end{table}

The NLTE effects are significant for several species in B-star atmospheres. Corrections for He\,{\sc i} and Si\,{\sc ii/iii} are important due to photoionization and radiative pumping \citep[e.g.,][]{auer1973analyses}. Sulfur (S\,{\sc ii/iii}) is susceptible to overpopulation via collisions and radiative transitions \citep[e.g.,][]{vrancken1996non}, while iron (Fe\,{\sc iii}) experiences overionization that can lead to abundance underestimates if neglected \citep[e.g.,][]{nieva2012present}. Therefore, detailed NLTE line formation calculations were conducted for all species reported here.

\begin{figure*}
    \centering
    \includegraphics[width=0.99\linewidth]{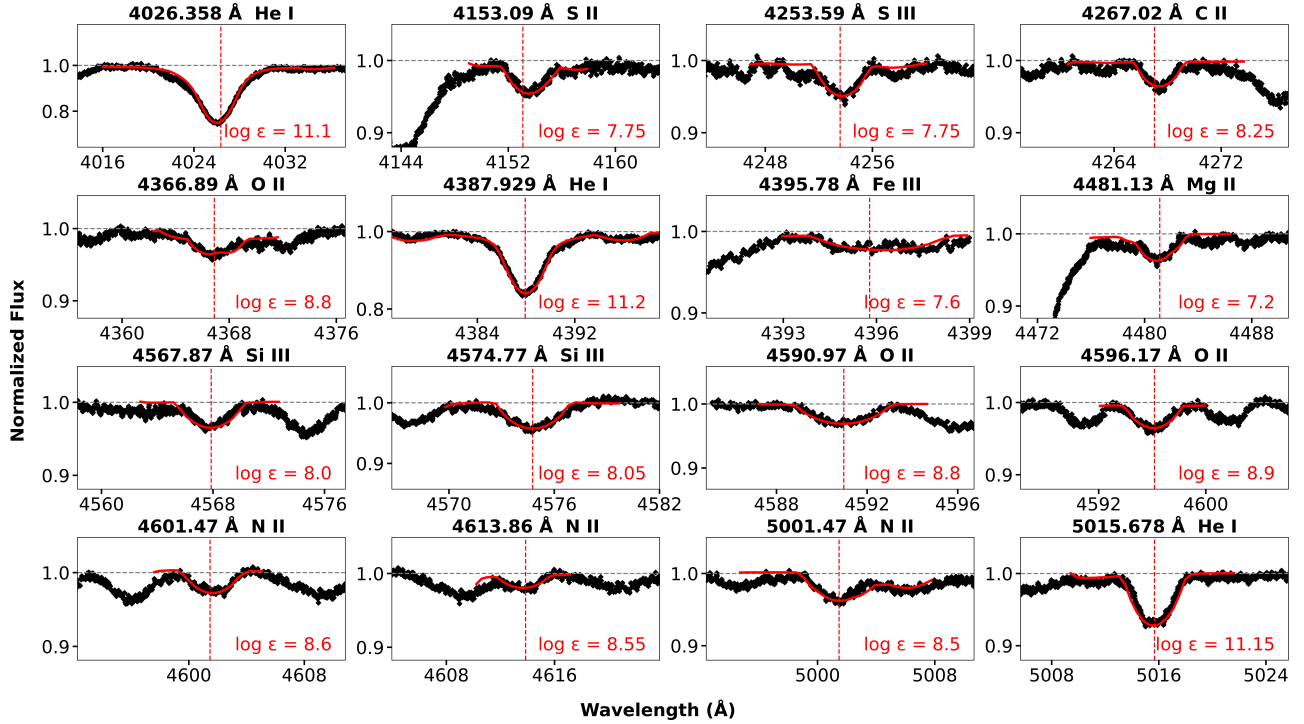}
    \caption{Selected line profile fits for the ET\,Cru primary component from FEROS spectra. Observed spectra are shown as black dots, with the best-fit synthetic spectra overlaid in red. The dashed horizontal line indicates the continuum level. The derived logarithmic abundance (log$\epsilon$) for each species is indicated in the corresponding panel.}
    \label{fig:primary_synth}
\end{figure*}

\begin{figure*}
    \centering
    \includegraphics[width=0.99\linewidth]{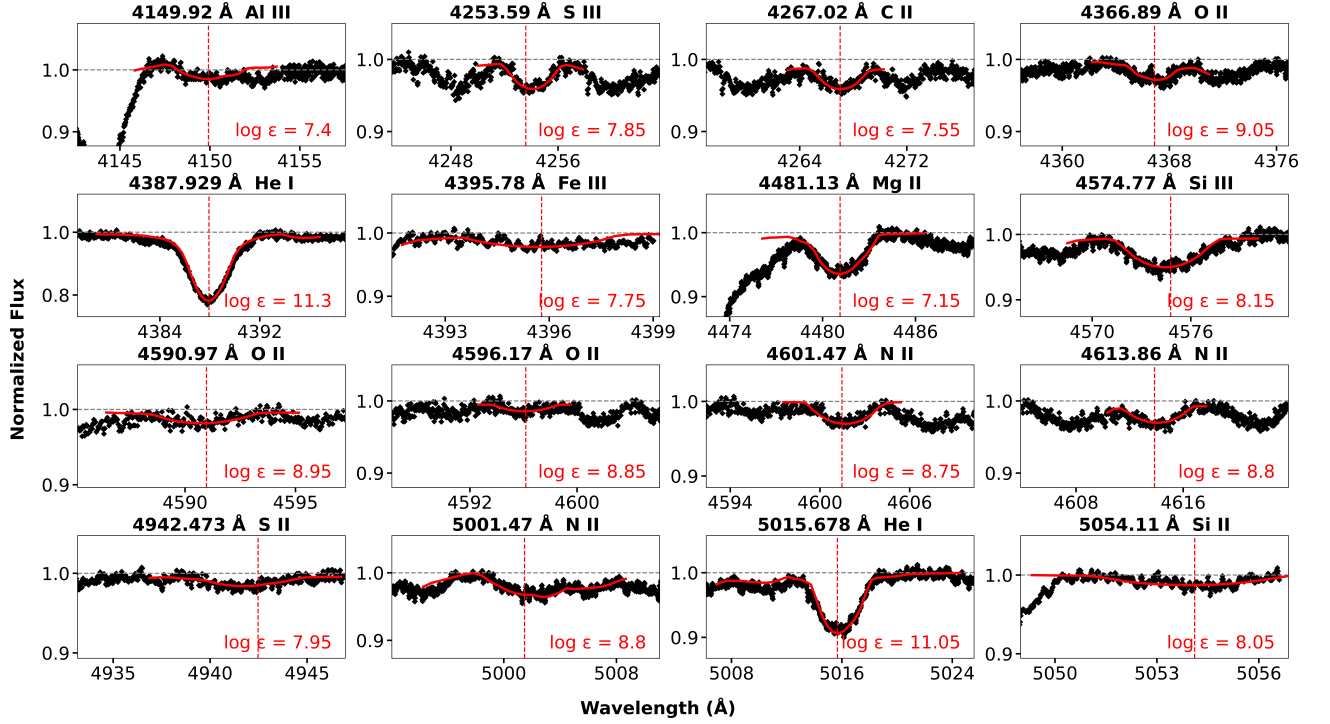}
    \caption{Same as Figure~\ref{fig:primary_synth}, but for the secondary component. Line profiles are shown for Al\,\textsc{iii} ($\lambda$4149.92\,\AA), S\,\textsc{iii} ($\lambda$4253.59\,\AA), C\,\textsc{ii} ($\lambda$4267.02\,\AA), O\,\textsc{ii} ($\lambda$4366.89, 4590.97, 4596.17\,\AA), He\,\textsc{i} ($\lambda$4387.929, 5015.678\,\AA), Fe\,\textsc{iii} ($\lambda$4395.78\,\AA), Mg\,\textsc{ii} ($\lambda$4481.13\,\AA), Si\,\textsc{iii} ($\lambda$4574.77\,\AA), N\,\textsc{ii} ($\lambda$4601.47, 4613.86, 5001.47\,\AA), and S\,\textsc{ii} ($\lambda$4942.473\,\AA). The derived logarithmic abundance ($\log \epsilon$) for each species is indicated in the corresponding panels.}
    \label{fig:secondary_synth}
\end{figure*}

\begin{table*}
	\setlength{\tabcolsep}{15pt}
	\renewcommand{\arraystretch}{1.05}
\centering
\caption{Multi-stellar parameters and heuristic errors of ET\,Cru.} \label{tab:parameters}
\begin{tabular}{lccc}\toprule
Parameter & Symbol  & Primary & Secondary \\
\midrule
Equatorial coordinate (Sexagesimal)             & $(\alpha, \delta)_{\rm J2000}$ & \multicolumn{2}{c}{12:50:28.04, -60:39:48.81} \\
Galactic coordinate (Decimal)                   & $(l, b)_{\rm J2000}$  & \multicolumn{2}{c}{302.812957, 2.207965} \\
Ephemerides time (d)                            & \emph{T}$_{\rm 0}$    & \multicolumn{2}{c}{$2459321.8934_{-0.0002}^{+0.0002}$} \\ 
Orbital period (d)                              & \emph{P}              & \multicolumn{2}{c}{$2.04386211_{-0.00000039}^{+0.00000039}$}  \\ 
Separation ($R_\odot$)                          & \emph{a}              & \multicolumn{2}{c}{$18.24^{+0.07}_{-0.06}$} \\ 
Combined magnitude (mag)$^{1,*}$             & $m_{\rm V, G, G_{BP}, G_{RP}}$  & \multicolumn{2}{c}{$9.209_{-0.017}^{+0.017}$, $9.045_{-0.015}^{+0.015}$, $9.082_{-0.015}^{+0.015}$, $8.908_{-0.015}^{+0.015}$}   \\
Combined color index (mag)                      & $B-V$                 & \multicolumn{2}{c}{$0.060_{-0.039}^{+0.035}$}          \\
Color excess (mag)                              & $E(B-V)$              & \multicolumn{2}{c}{$0.36_{-0.03}^{+0.03}$}                \\
Interstellar Extinction (mag)                   & $A_{\rm V, G, G_{BP}, G_{RP}}$    & \multicolumn{2}{c}{$1.113, 1.158, 1.362, 0.704$}     \\
Systemic velocity (km\,s$^{-1}$)                & $v_{\gamma}$          & \multicolumn{2}{c}{$-3.9_{-1.8}^{+1.9}$}        \\
Orbital inclination ($^{\circ}$)                & \emph{i}              & \multicolumn{2}{c}{$70.78_{-0.12}^{+0.11}$}      \\
Mass ratio                                      & \emph{q}              & \multicolumn{2}{c}{$0.4476_{-0.0036}^{+0.0046}$}   \\
Eccentricity                                    & \emph{e}              & \multicolumn{2}{c}{$0$ (fixed)}  \\
Argument of periastron (rad)                       & \emph{w}              & \multicolumn{2}{c}{$0$ (fixed)}    \\
Spectral type                                   & Sp                    & B0-V & B2-III\\
Metallicity (dex)                              & [Fe/H] & 0.20 $\pm$ 0.15  & 0.20 $\pm$ 0.10                \\
Mass ($M_\odot$)                                & \emph{M}              & $13.408_{-0.153}^{+0.135}$  & $6.001_{-0.082}^{+0.089}$ \\
Radius ($R_\odot$)                              & \emph{R}              & $5.575_{-0.021}^{+0.020}$  & $5.676_{-0.027}^{+0.030}$ \\
Surface gravity (cgs)                           & $\log g$              & $4.073_{-0.008}^{+0.008}$  & $3.708_{-0.010}^{+0.011}$ \\
Temperature (K)                                 & $T_{\rm eff}$         & $29000^{+1000}_{-1000}$     & $20200^{+1000}_{-1000}$ \\
Light ratio (ASAS-\textit{V})                        & $l/l_{\rm{total}}$ &  $0.595_{-0.010}^{+0.012}$  & $0.404_{-0.012}^{+0.010}$ \\
Light ratio (\textit{\textit{Gaia} G})                        & $l/l_{\rm{total}}$    & $0.6108_{-0.0005}^{+0.0006}$     & $0.3807_{-0.0006}^{+0.0005}$  \\
Spot relative temperature                            &  Temp.  &  $0.833_{-0.055}^{+0.035}$ &  \\
Spot colatitude ($^{\circ}$)  & $\theta$  & $106.2_{-6.1}^{+7.0}$ & \\
Spot latitude ($^{\circ}$)    & $\phi$    & $12.0_{-2.3}^{+2.1}$ & \\     
Spot radius ($^{\circ}$)    & $r$ &  $25.1_{-3.8}^{+3.8}$ &\\ 
Luminosity ($L_\odot$)                          & $\log$ \emph{L}       & $4.297_{-0.064}^{+0.062}$     & $3.684_{-0.092}^{+0.089}$ \\
\emph{V} magnitude (mag)             & $m_{\rm V\,1,2}$      & $9.771^{+0.018}_{-0.022}$     & $10.193^{+0.033}_{-0.027}$  \\
\emph{G} magnitude (mag)             & $m_{\rm G\,1,2}$      & $9.580_{-0.016}^{+0.016}$     & $10.094_{-0.017}^{+0.016}$  \\
$G_{\rm BP}$ magnitude (mag)         & $m_{\rm G_{BP}\,1,2}$      & $9.599_{-0.016}^{+0.016}$     & $10.135_{-0.017}^{+0.016}$ \\
$G_{\rm RP}$ magnitude (mag)         & $m_{\rm G_{RP}\,1,2}$      & $9.472_{-0.016}^{+0.016}$     & $9.889_{-0.017}^{+0.016}$  \\
Bolometric magnitude (mag)                      & $M_{\rm Bol}$         & $-6.002_{-0.155}^{+0.160}$     & $-4.470_{-0.222}^{+0.230}$ \\
Bolometric correction (mag)$^{2,3}$             & \emph{BC$_{\rm V}$}   & $-2.451^{+0.095}_{-0.093}$     & $-1.533^{+0.115}_{-0.118}$ \\
Bolometric correction (mag)$^{4}$             & \emph{BC$_{\rm G}$}   & $-2.406_{-0.090}^{+0.088}$     & $-1.508_{-0.113}^{+0.116}$ \\
Bolometric correction (mag)$^{4}$             & \emph{BC$_{\rm G_{BP}}$}   & $-2.171_{-0.086}^{+0.093}$     & $-1.326_{-0.095}^{+0.108}$ \\
Bolometric correction (mag)$^{4}$             & \emph{BC$_{\rm G_{RP}}$}   & $-2.763_{-0.117}^{+0.100}$     & $-1.776_{-0.127}^{+0.132}$ \\
Absolute \emph{V} magnitude (mag)    & $M_{\rm V\,1,2}$      & $-3.551_{-0.251}^{+0.254}$     & $-2.937_{-0.336}^{+0.348}$  \\
Absolute \emph{G} magnitude (mag)    & $M_{\rm G\,1,2}$      & $-3.596_{-0.247}^{+0.150}$     & $-2.962_{-0.337}^{+0.344}$  \\
Absolute $G_{\rm RP}$ magnitude (mag) & $M_{\rm G_{BP}\,1,2}$ & $-3.831_{-0.233}^{+0.247}$     & $-3.145_{-0.329}^{+0.335}$  \\
Absolute $G_{\rm RP}$ magnitude (mag) & $M_{\rm G_{RP}\,1,2}$ & $-3.239_{-0.256}^{+0.338}$     & $-2.695_{-0.335}^{+0.357}$  \\
Computed synchronization velocity (km\,s$^{-1}$)& $v_{\rm synch}$       & $131.1^{+2.3}_{-2.3}$     & $141.6^{+3.3}_{-3.3}$ \\
Calculated rotation velocity (km\,s$^{-1}$)     & $v_{\rm rot}$         & $145^{+12}_{-12}$       & $150^{+2}_{-2}$ \\
Photometric distance (pc)$^{3}$                       &  \emph{d}             & $2656_{-125}^{+373}$ & $2580_{-103}^{+403}$     \\
{\it Gaia} DR3 distance (pc)$^5$                        & $d_{\,\varpi}$    & \multicolumn{2}{c}{$1869_{-54}^{+58}$}         \\
\bottomrule
\multicolumn{4}{l}{\small $^1$\cite{Hog2000}, $^2$\cite{Yucel2026}, $^3$\cite{Johnson1953}, $^4$\cite{Eker2026}} \\
\multicolumn{4}{l}{$^5$\cite{Gaia2023}} \\
\multicolumn{4}{l}{$*$ Calculated from the maxima of the light curves} \\
\label{table:fund}
\end{tabular}
\end{table*}

The final photospheric elemental abundances for both components of ET\,Cru are presented in Table~\ref{tab:etcru_abundances} and are compared visually in Figure~\ref{fig:abund_pattern}. The derived logarithmic abundances, \(\log \epsilon(X)\), reveal distinct chemical patterns that elucidate the system's evolutionary history, which is dominated by mass transfer.

The secondary component (the present-day donor) exhibits classic signatures of CNO-processed material: strong nitrogen enrichment (\(\log \epsilon(\mathrm{N\,\textsc{ii}}) = 8.87 \pm 0.08\) dex), significant carbon depletion (\(\log \epsilon(\mathrm{C\,\textsc{ii}}) = 7.55\) dex), and helium (\((\log \epsilon(\mathrm{He}\,\textsc{i}) = 11.22 \pm 0.11\) dex) comparable to the primary. This pattern indicates that envelope stripping has exposed the star's chemically processed interior, consistent with its role as the original, more massive star in the Algol paradigm. Aluminum is detected only in the secondary. Other elements show more subtle variations: sulfur abundances (S\,{\sc ii}, S\,{\sc iii}) are slightly enhanced in the secondary by $\approx$0.2 dex, while silicon (Si\,{\sc iii}) is nearly identical in both stars. Magnesium (Mg\,{\sc iii}) is marginally depleted in the secondary by $\approx$0.05 dex.

The primary component (the gainer) displays more subtle but distinct signatures of pollution by the transferred material, including moderate nitrogen enrichment (\(\log \epsilon(\mathrm{N}\,\textsc{iii}) = 8.60 \pm 0.11\) dex) and carbon depletion (\(\log \epsilon(\mathrm{C}\,\textsc{ii}) = 8.25\) dex). Oxygen abundances are similar in both stars and consistent with near-solar values. The abundances of other metals, as well as iron, are also near-solar in both components, reflecting either the system's initial metallicity or the effects of shallow mixing in the gainer's outer layers.

Quantifying the chemical alteration using solar abundances from \citet{asplund2009chemical}, the primary shows [C/H] = $-0.18$ dex, [N/H] = $+0.77$ dex, and [C/N] = $-0.95$ dex (C/N\footnote{The ratio of carbon and nitrogen has been demonstrated to be a highly sensitive indicator of hydrogen-core CNO nucleosynthesis. Furthermore, it has been observed that this ratio can vary in response to distinct mixing processes that occur prior to and following a mass transfer episode.}$\approx$0.45). The secondary displays extreme values: [C/H] = $-0.88$ dex, [N/H] = $+1.04$ dex, and [C/N] = $-1.92$ dex (C/N$\approx$0.05). This exceptionally low C/N ratio confirms that the secondary has been stripped down to layers that underwent thorough CNO cycle processing.

The quality of the spectral synthesis was verified through detailed line profile fitting. Examples of the agreement between the observed and synthetic spectra for both components are presented in Figures ~\ref{fig:primary_synth} and \ref{fig:secondary_synth}, demonstrating the reliability of our atmospheric parameters and abundance determination.

%------------------------------------------------------------------
\subsection{Astrophysical Parameters and Distance}\label{sec:params}

A reliable determination of the astrophysical parameters of the components in an eclipsing binary system is essential to understand their evolutionary status and place them within the broader context of stellar astrophysics. Once orbital and photometric solutions are established, fundamental parameters such as mass, radius, effective temperature, and luminosity can be derived with high precision. These quantities allow for a direct comparison with stellar evolutionary models and provide constraints on processes such as mass transfer and angular momentum loss. In addition, the distance to the system can be estimated by combining photometric and spectroscopic information with bolometric corrections, offering an independent check against astrometric measurements (e.g., \textit{Gaia}). In Table~\ref{tab:parameters}, we present the astrophysical parameters of ET\,Cru as derived from our analysis, and we subsequently discuss its distance in comparison with values reported in available catalogs.

\textit{Gaia} DR3 reports the parallax of ET\,Cru as $\varpi=0.5351\pm0.016$\,mas, corresponding to a distance of $d=1869_{-30}^{+31}$\,pc. However, using the combined Bessel-$V$-magnitude ($9.2^{m}$),  and the light ratios of the components given in Table~\ref{tab:solution}, the visual magnitude of each component have been calculated as $9.771^{+0.018}_{-0.022}$ mag and $10.193^{+0.033}_{-0.027}$ mag for the primary and secondary components, respectively. Later on, bolometric corrections ($BC$) for each component were calculated by using the $BC-T_{\rm eff}$ formula given by \cite{Yucel2026} with the calculated spectroscopic temperatures of each component as $-2.451^{+0.095}_{-0.093}$ mag and $-1.533^{+0.115}_{-0.118}$ mag, which subsequently correspond to absolute $V$-magnitudes ($M_{\rm V}$) as $-3.444^{+0.308}_{-0.306}$ mag and $-2.959^{+0.354}_{-0.358}$ mag for the primary and secondary components, respectively. As a result, by using distance modulus, which enabled us to calculate, by using interstellar extinction ($A_{\rm V}=1.11^{m}$), the distance of each component is estimated to be $2767^{+308}_{-285}$\,pc (for the primary component) and $2534^{+387}_{-342}$\,pc (for the secondary component).

In addition to the Johnson $V$-band, we also derived photometric distances using the \textit{Gaia} $G$, $G_{\rm BP}$, and $G_{\rm RP}$ bands. For the primary component, the resulting distances are $2523\pm300$\,pc ($G$), $2593\pm250$\,pc ($G_{\rm BP}$), and $2521\pm280$\,pc ($G_{\rm RP}$), while for the secondary component they are $2500\pm350$\,pc ($G$), $2420\pm330$\,pc ($G_{\rm BP}$), and $2378\pm340$\,pc ($G_{\rm RP}$). All photometric estimates consistently place the system at a distance of $\sim$2.4--2.7\,kpc.

These values deviate from the \textit{Gaia} DR3-based distance, $d=1869_{-30}^{+31}$\,pc, derived from the parallax $\varpi=0.5351\pm0.016$\,mas. On the other hand, the \textit{Gaia} DR2-based distance of ET\,Cru was reported as $2442.0_{-198}^{+236}$\,pc by \citet{BailerJones2018}, which is closer to our photometric estimates.

\begin{figure*}
    \centering
    \includegraphics[width=0.99\linewidth]{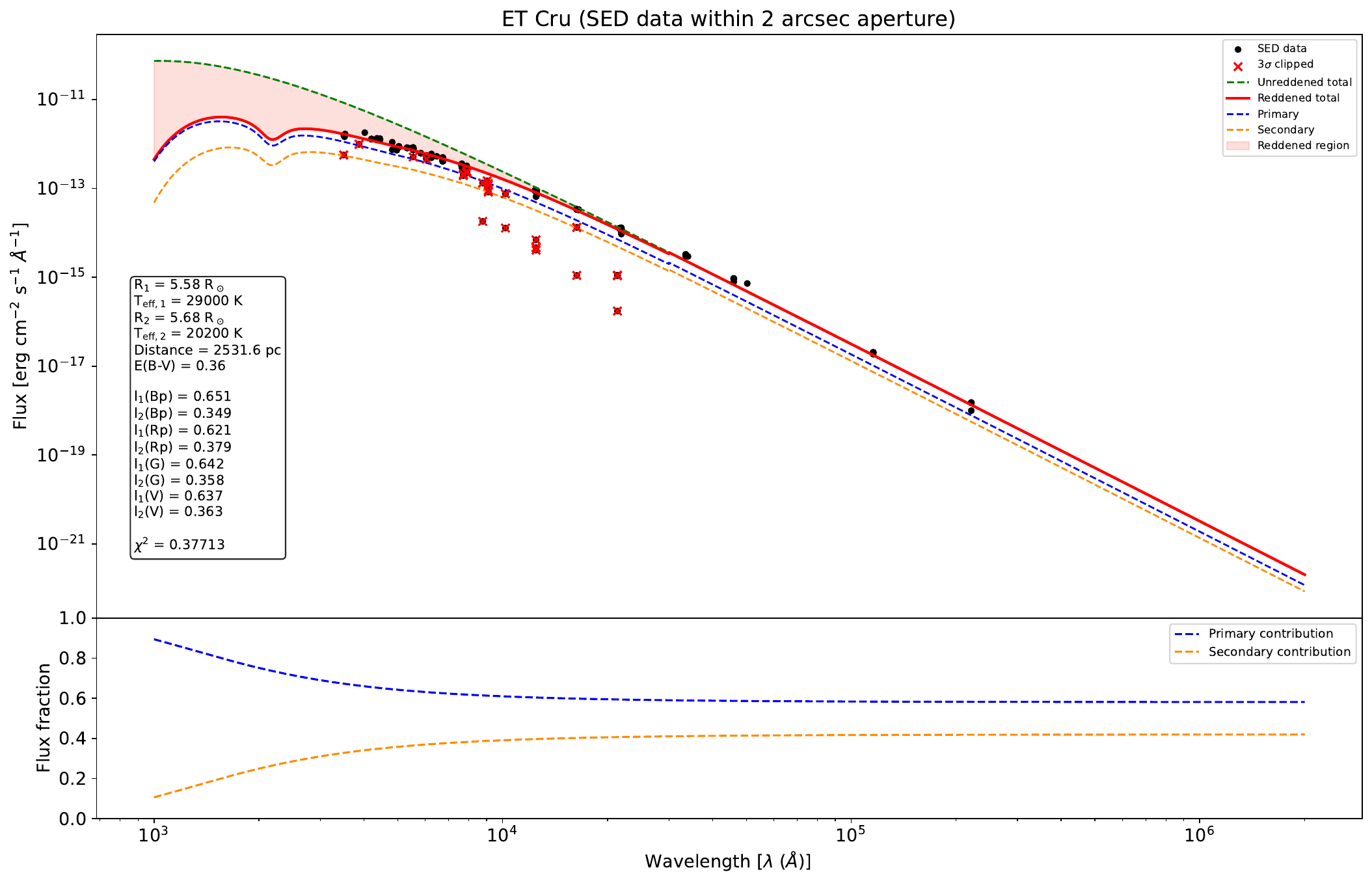}
    \caption{SED of ET\,Cru constructed from multi-wavelength photometric data. Black circles represent the observed fluxes, and red crosses indicate data points excluded by the $3\sigma$ clipping procedure. The solid red curve shows the total reddened model SED, while the dashed green curve corresponds to the intrinsic (unreddened) SED. The dashed blue and orange curves illustrate the individual flux contributions of the primary and secondary components, respectively. The model adopts a single distance of $\sim$2.5\,kpc, which provides a significantly better agreement with the observed SED than the distance inferred from the \textit{Gaia} parallax.}
    \label{fig:SED}
\end{figure*}

To further investigate this discrepancy, we analyzed the multi-wavelength spectral energy distribution (SED) of ET\,Cru using flux data from the VizieR database, following the methodology described by \citet{Bakis2022} and \citet{Eker2023}. In Figure~\ref{fig:SED}, we present the SED model constructed using the temperatures and radii of the components (Table~\ref{table:fund}), together with the observed broadband photometric data. As shown in Figure~\ref{fig:SED}, a distance of $\sim$2.5\,kpc provides an excellent fit to the observed SED.

Although the \textit{Gaia} DR3 astrometric solution appears reliable (RUWE = 0.814), the systematic offset between the parallax-based distance and the photometric estimates suggests that additional factors, such as binary effects or unresolved systematics, may influence the astrometric measurement.

%------------------------------------------------------------------

\section{DISCUSSION}
\label{sec:discussion}

In this study, by combining photometric and spectroscopic data, we conducted a comprehensive analysis of the semi-detached binary system ET\,Cru and derived the fundamental parameters of each component with high precision. For the first time in the literature, spectral disentangling techniques were employed to obtain accurate effective temperatures and chemical abundances—up to nine elements—for both components. Our analysis indicates that the masses of the primary and secondary are $13.408_{-0.153}^{+0.135}$ and $6.001_{-0.082}^{+0.089}\,M_\odot$, respectively. Similarly, the radii of the components are $5.575_{-0.021}^{+0.020}$ and $5.676_{-0.027}^{+0.027}\,R_\odot$ for the primary and secondary, respectively. The fundamental parameters of ET\,Cru offer profound insights into the evolutionary history of massive interacting binaries. The system provides a clear example of the Algol paradox, wherein the currently less massive secondary (\(6.00 \pm 0.08\,M_{\odot}\), Table~\ref{table:fund}) is the more evolved component, as evidenced by its larger radius (\(5.68 \pm 0.01\,R_{\odot}\)), lower surface gravity (\(\log g \approx 3.71\)). This confirms that the secondary was originally the more massive star and has undergone extensive mass loss via Roche-lobe overflow \citep{Hilditch2001}.

An important outcome of our analysis concerns the discrepancy between the photometric/spectroscopic distance estimates and the \textit{Gaia} parallax. While \textit{Gaia} DR3 reports a distance of $1869^{+31}_{-30}$\,pc, our independent determinations based on the combined $V$-magnitude, extinction, and light contributions of the components yield values between $\sim$2.4--2.7\,kpc, supported by consistent results obtained from the \textit{Gaia} $G$, $G_{\rm BP}$, and $G_{\rm RP}$ bands, consistent with the multi-wavelength SED modelling. 

The consistency between independent photometric passbands and the SED-based distance, which provides an excellent fit to the observed flux distribution, supports a larger distance estimate for the system.

Although the \textit{Gaia} DR3 astrometric solution appears reliable (RUWE = 0.814), the systematic offset between the parallax-based distance and the photometric estimates may indicate the presence of additional effects influencing the astrometric solution.

In interacting binary systems such as ET\,Cru, factors such as orbital motion, light asymmetries, and possible photocenter shifts can affect the measured parallax. The derived distance also places ET\,Cru within the regions of OB associations associated with the Crux constellation, although its galactic position lies slightly outside the boundaries defined by \citet{Tovmassian1996}. 

ET\,Cru therefore represents a system where independent photometric and SED-based analyses provide a complementary perspective to the astrometric distance, highlighting the need for caution when interpreting parallaxes of interacting binaries.

The abundance patterns of ET\,Cru fit coherently within the theoretical framework for post-mass-transfer Algol binaries, while standing out due to the severity of the donor's chemical anomalies. Comparisons with other systems underscore their distinct evolutionary stages.

The primary component of u~Her shows a C/N ratio of $\sim0.9$ \citep{Kolbas2014}, while the primaries of $\delta$~Lib and the prototype Algol ($\beta$~Per) exhibit ratios of $\sim1.5$ and $\sim2.0$, respectively---all significantly higher than the extreme ratio found in ET\,Cru's secondary. Historical studies of systems such as U~Cep and U~Sge have documented nitrogen enrichment and carbon depletion in their secondaries, but the anomalies ([C/Fe] $\approx -0.5$\,dex, [N/Fe] $\approx +0.5$\,dex) were less pronounced \citep{parthasarathy1983chemical}.

A broader survey by \citet{Ibanoglu2012} of 18 Algol primaries found an average carbon depletion of [C/H] $\approx -0.8$\,dex, with nitrogen typically near solar levels. This suggests that accretor pollution in classical Algols is usually moderate. The stark contrast with ET\,Cru's secondary implies that this system has experienced more extensive envelope stripping.

In this context, ET\,Cru serves as a pivotal case. Its secondary offers unambiguous spectroscopic evidence of deep nucleosynthetic exposure, whereas the primary's composition confirms its role as a polluted gainer. Together, these findings establish ET\,Cru not merely as a textbook Algol system, but as a particularly evolved and chemically illustrative benchmark for understanding advanced binary interactions and stellar evolution.

When compared with other Algol-type binaries in the literature, ET\,Cru fits well within the established framework but also distinguishes itself by the severity of its abundance anomalies. For instance, \cite{Kolbas2014} reported a C/N ratio of $\approx$0.9 for the primary of u Her, while \cite{Dervisoglu2018} found a value of $\approx$1.5 ($\approx$0.4) for $\delta$ Lib, and \cite{Kolbas2015} derived $\approx$2.0 ($\pm$0.4), half of the solar value of C/N$_{\rm \odot}$=4.0$\pm$0.7, for the primary of the prototype system Algol ($\beta$ Per, HD\,19356)—all indicating milder chemical processing compared to ET\,Cru. \cite{Parthasarath1983}, \cite{Cugier1988}, \cite{Cugier1989}, \cite{Tomkin1993} and \cite{Ibanoglu2012} detected C depletion in Algol and in Algol-type systems. 

\cite{Parthasarath1983} documented nitrogen enhancement and carbon depletion in the secondaries of U Cep and U Sge ([C/Fe]$\approx$-0.5 dex and [N/Fe]$\approx$+0.5 dex), though not as extreme as seen in ET\,Cru’s secondary. 

A broader survey by \cite{Ibanoglu2012}, covering 18 Algol primaries, confirmed widespread carbon depletion with an average [C/H] $\approx$ -0.8 dex (relative to the Sun), while nitrogen abundances remained near solar, indicating that the accreting primaries in most systems undergo only moderate surface pollution. Their spectroscopic analysis was based on one ionized C line at 4267 \AA. This average contrasts strongly with the more severe [C/N] inversion seen in ET\,Cru, particularly in the donor component, suggesting a deeper level of mass loss and envelope stripping in this system than is typical for classical Algol binaries. 

For instance, \cite{Tomkin1993} reported carbon abundances for eight Algol primaries, with [C/H] values ranging from -0.62 to +0.24 dex (see their Table 3). In contrast, the secondary component of ET\,Cru exhibits [C/H] $\approx$ –0.88 dex, which is more depleted than most of those systems, while its [N/H] $\approx$ +1.04 dex far exceeds typical enhancements. This severe C/N inversion underscores that ET\,Cru’s donor has been stripped down to layers that underwent thorough CNO-cycle processing, making it one of the most chemically extreme Algol donors spectroscopically documented.

In this context, the secondary component of ET\,Cru emerges as an exceptional case, offering clear spectroscopic evidence of deep CNO-cycle exposure. Simultaneously, the altered surface composition of the primary confirms its role as a rejuvenated mass gainer. Together, these findings establish ET\,Cru as a textbook example of classical Algol evolution and a particularly chemically evolved and dynamically illustrative system within its class.

In the broader landscape of massive binary evolution, ET\,Cru serves as a benchmark for understanding late-stage mass transfer and envelope stripping. Its chemical extremity suggests that binary interaction can expose deeper layers of nucleosynthesis than typically observed in classical Algols, offering a unique window into the advanced phases of massive star evolution before systems evolve into compact binaries or supernova progenitors \citep{Eldridge2017}. Future binary population synthesis studies may use systems like ET\,Cru to calibrate mass-transfer efficiency and mixing processes in high-mass binaries.

\begin{acknowledgments}

 We thank the anonymous referee for their insightful and constructive suggestions that significantly improved the paper. We thank T\"{U}B\.{I}TAK for funding this research under project number 109T449. This work is partly based on observations collected with the FEROS spectrograph available at the European Organisation for Astronomical Research in the Southern Hemisphere under ESO programme 086.D-0236. The numerical calculations reported in this paper were partially performed at T\"{U}B\.{I}TAK ULAKB\.{I}M, High Performance and Grid Computing Center (TRUBA resources). This research made use of the Astrophysics Data System, funded by NASA under Cooperative Agreement 80NSSC25M7105. The VizieR and Simbad databases at CDS, Strasbourg, France were invaluable for the project as were data from the European Space Agency (ESA) mission \emph{Gaia}\footnote{https://www.cosmos.esa.int/gaia}, processed by the \emph{Gaia} Data Processing and Analysis Consortium (DPAC)\footnote{https://www.cosmos.esa.int/web/gaia/dpac/consortium}. Funding for the DPAC was provided by national institutions, in particular, those participating in the \emph{Gaia} Multilateral Agreement. We acknowledge the use of TESS High Level Science Products (HLSP) produced by the Quick-Look Pipeline (QLP) at the TESS Science Office at MIT, which are publicly available from the Mikulski Archive for Space Telescopes (MAST). Funding for the TESS mission is provided by NASA's Science Mission directorate. The data described here may be obtained from the MAST archive via \cite{tessdata}. This research made use of Lightkurve, a Python package for Kepler and TESS data analysis \citep{lk}.

\software{
\texttt{Astropy} \citep{astro1,astro2,astro3}, 
\texttt{corner} \citep{corner}, 
\texttt{lightkurve} \citep{lk},
\texttt{Matplotlib} \citep{matplotlib},
\texttt{NumPy} \citep{numpy}, 
\texttt{PHOEBE} \citep{phoebe2_2016, phoebe2_2018, phoebe2_2020, phoebe2_2020b},
\texttt{SciPy} \citep{scipy},
\texttt{SYNSPEC} \citep{Hubeny1995} , and
\texttt{TLUSTY} \citep{Hubeny1995}
}

\end{acknowledgments}
\appendix
\renewcommand{\thefigure}{A\arabic{figure}}
\setcounter{figure}{0}

\section{\textit{TESS} Light Curves}

\begin{figure}[h]
    \centering
    \includegraphics[width=0.95\linewidth]{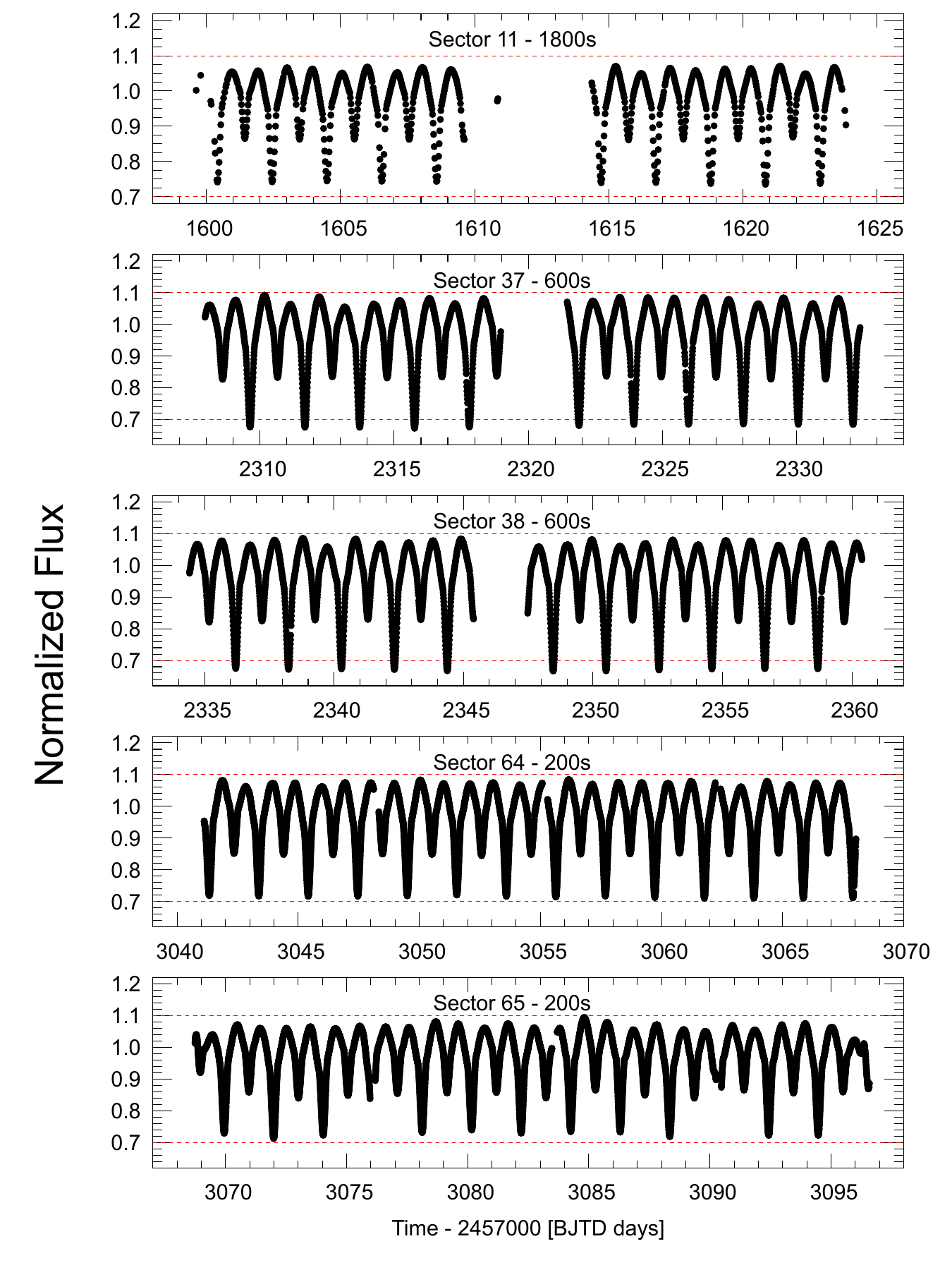}
    \caption{\textit{TESS} light curves of ET\,Cru obtained in five sectors (11, 37, 38, 64, and 65), with exposure times of 1800\,s, 600\,s, 600\,s, 200\,s, and 200\,s from top to bottom, respectively. Only data points passing the \texttt{quality\_mask='hard'} criterion are shown. The light curves exhibit noticeable variations in overall flux levels and amplitudes between different sectors and cadences, particularly outside eclipse phases, suggesting the presence of systematic effects.}
    \label{fig:TESS}
\end{figure}

The figure~\ref{fig:TESS} shows the TESS light curves of ET\,Cru obtained in five sectors (more details are given in the corresponding caption).

\section{Hyrdogen-Helium Spectral Lines}

\begin{figure}[h]
    \centering
    \includegraphics[width=0.49\linewidth]{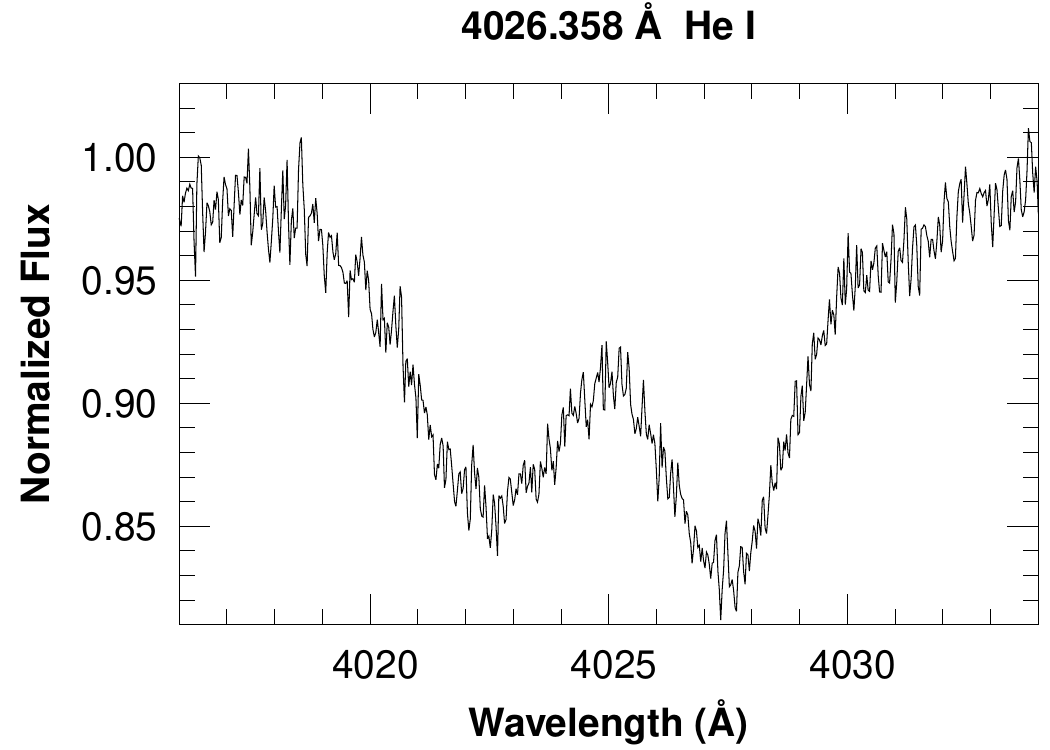}
    \includegraphics[width=0.49\linewidth]{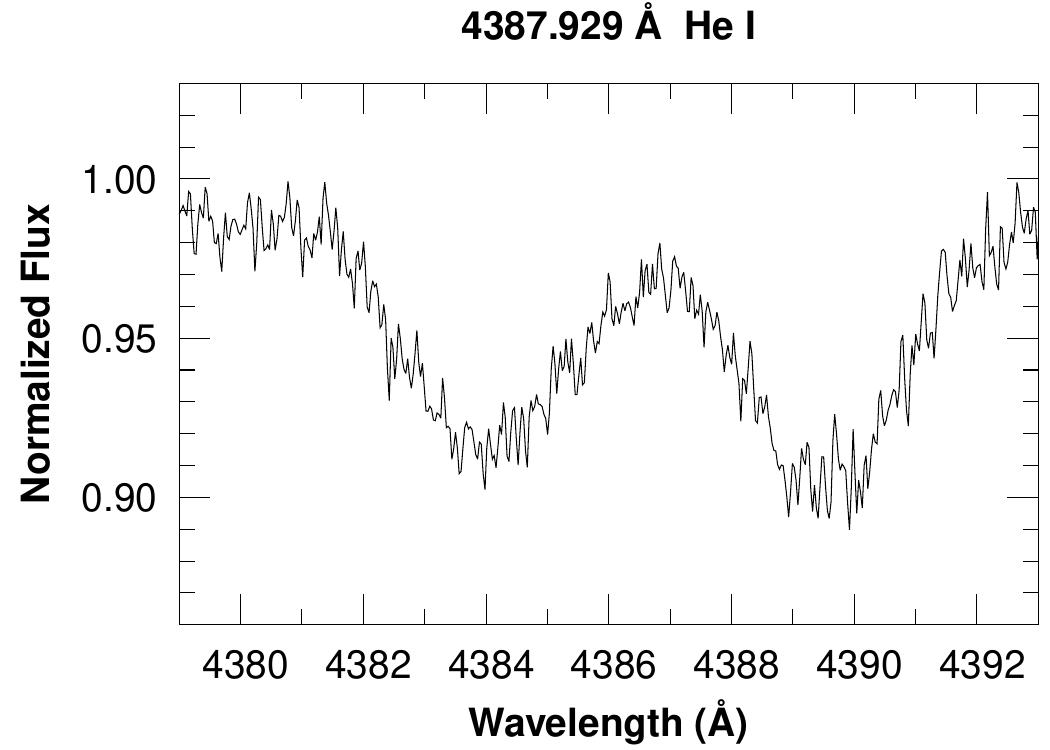} \\
    \vspace{1cm}
    \includegraphics[width=0.49\linewidth]{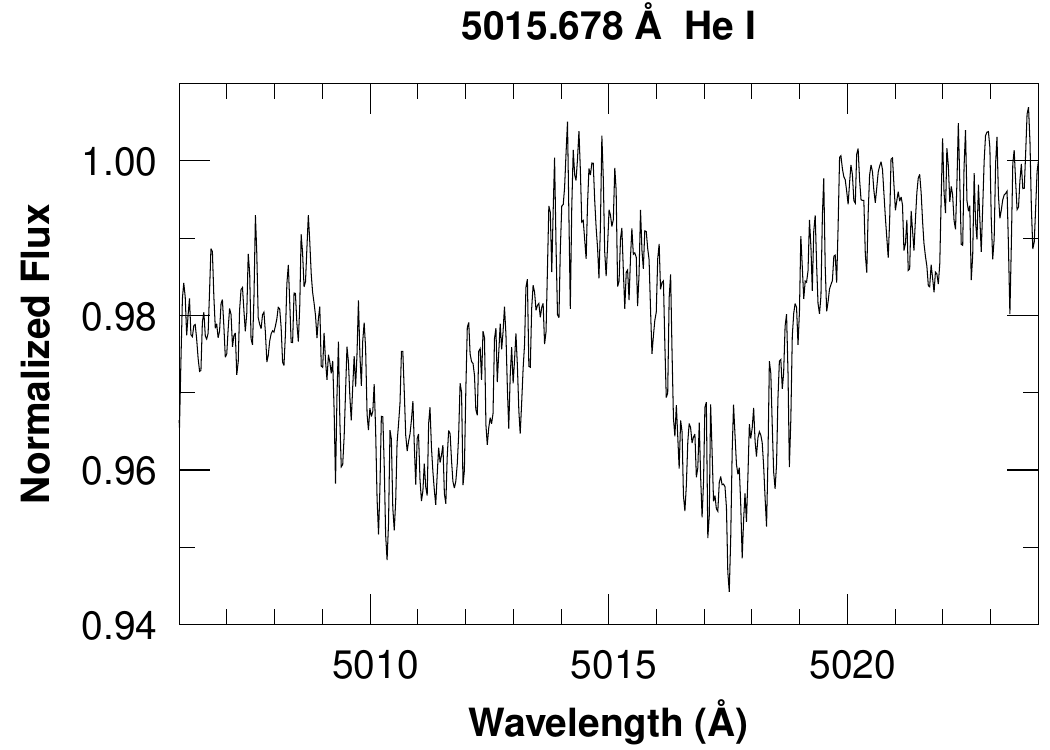}
    \includegraphics[width=0.49\linewidth]{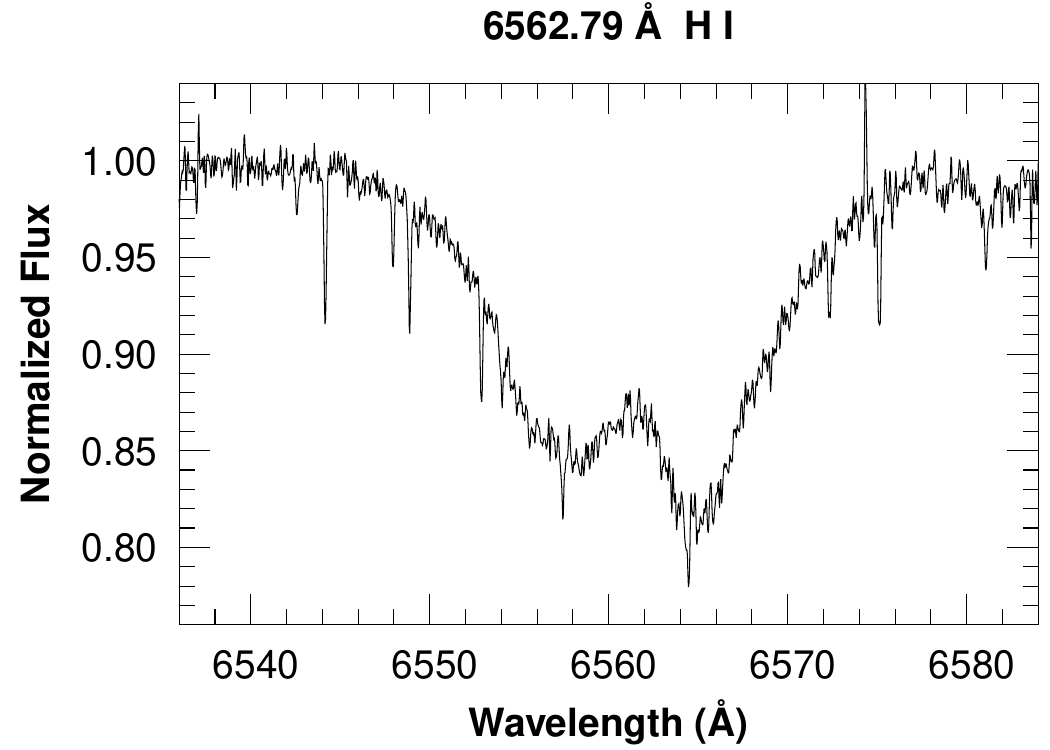}
    \caption{H$\alpha$ and HeI lines at where the spot visibility is maximum and at the closest orbital phase ($\phi=0.819$) among our spectral dataset.}
    \label{fig:emission}
\end{figure}

The figure~\ref{fig:emission} shows the various helium and hydrogen lines from the spectrum that have been obtained at $\phi=0.819$ where the spot is directly in the line of sight.

\section{Table of SED Data of ET~Cru}

\begin{tabular}{rcccc}
\hline
Number & Wavelength ($\mathrm{\AA}$) & Flux (Jy) & Error (Jy) & Survey:Filter \\
\hline
1&3502.63&0.631& - & Cousins:U \\
2&3521.46&0.601& - & SDSS:u \\
3&3533.44&0.701&0.01& Johnson:U \\
4&4022.8&0.972& - & HIP:Hp \\
5&4205.92&0.754&0.011& HIP:BT \\
6&4363&0.852& - & Cousins:B \\
7&4445.1&0.819&0.015& Johnson:B \\
8&4823.31&0.576&0.039& SDSS:g \\
9&4971.83&0.656&0.068& SkyMapper/SkyMapper:g \\
10&5039.48&0.75&0.009& GAIA/GAIA3:Gbp \\
11&5049.65&0.725&0.012& GAIA/GAIA2:Gbp \\
12&5322.64&0.774&0.015& HIP:VT \\
13&5473.45&0.809& - & Cousins:V \\
14&5540.88&0.755&0.002& Johnson:V \\
15&5826.37&0.702&0.004& GAIA/GAIA3:G \\
16&6044.73&0.588&0.03& SkyMapper/SkyMapper:r \\
17&6230.53&0.699&0.004& GAIA/GAIA2:G \\
18&6251.3&0.638&0.028& SDSS:r \\
19&6473.47&0.738& - & Cousins:R \\
20&6734.61&0.646&0.005& Gaia:G \\
21&7625.83&0.637&0.008& GAIA/GAIA3:Grp \\
22&7640.2&0.568&0.005& SDSS:i \\
23&7729.97&0.643&0.01& GAIA/GAIA2:Grp \\
24&7891.41&0.524&0.01& Cousins:I \\
25&12398.74&0.457&0.01& 2MASS:J \\
26&12472.46&0.339&0& VISTA:J \\
27&12508.86&0.462&0.001& Johnson:J \\
28&16311.44&0.301&0.012& Johnson:H \\
29&16506.19&0.302&0.018& 2MASS:H \\
30&21652.83&0.197&0.004& 2MASS:Ks \\
31&21915.41&0.193&0.001& Johnson:K \\
\hline
\multicolumn{5}{c}{\textbf{3$\sigma$-clipped data}} \\
\hline
1&3500.58&0.231&0.009& SkyMapper/SkyMapper:u \\
2&3873.47&0.494&0.01& SkyMapper/SkyMapper:v \\
3&5540.88&0.521&-& Johnson:V \\
4&6044.73&0.557&0.025& SkyMapper/SkyMapper:r \\
5&7718.03&0.395&0.001& SkyMapper/SkyMapper:i \\
6&7891.41&0.464&0.017& Cousins:I \\
7&8768.08&0.339&0& VISTA:Z \\
8&9024.18&0.403&0.018& SDSS:z \\
9&9096.42&0.228&0.02& SkyMapper/SkyMapper:z \\
10&10190.91&0.263&0& VISTA:Y \\
11&12472.46&0.024&0& VISTA:J \\
12&16321.2&0.118&0& VISTA:H \\
13&21352.31&0.00264&1.00E-05& VISTA:Ks \\
\end{tabular}

\bibliography{sample701}{}
\bibliographystyle{aasjournalv7}



\end{document}